\documentclass[a4paper, twocolumn]{article}

\usepackage{tikz} 
\usetikzlibrary{3d, positioning}

\usepackage{graphicx}
\usepackage[hidelinks]{hyperref}
\usepackage{subcaption}
\usepackage{comment} 
\usepackage{amsmath}
\usepackage{amsfonts}
\usepackage{braket}
\usepackage{bm}
\usepackage{mathtools}
\usepackage[title, titletoc]{appendix}
\usepackage{placeins}
\usepackage[T1]{fontenc}
\usepackage{lmodern}

\counterwithin{equation}{section}

\usepackage[
  backend=biber,
  style=numeric,
  sorting=none
]{biblatex}
\begin{document}

\twocolumn[{
\begin{center}
    \centering
    \vspace{0.3cm} 

    {\huge \bfseries Smart walkers in discrete space}

    \vspace{0.5cm} 

    {\large Gianluca Peri$^1$, Lorenzo Buffoni$^1$, 
    Giacomo Chiti$^1$, Duccio Fanelli$^1$
    \\\vspace{0.2cm}
    Raffaele Marino$^1$, Andrea Nocentini$^2$, Pier Paolo Panti$^2$,  \par}

    \vspace{1cm} 

    {\large
    $^1$Department of Physics and Astronomy, University of Florence \par
    {$^2$Sigma Lab, Tecnolink, Florence} \par
    }

    \vspace{0.5cm} 

    {\large \texttt{\{gianluca.peri, lorenzo.buffoni, raffaele.marino, duccio.fanelli\}@unifi.it} \par}
    \vspace{0.2cm}
    {\large {\texttt{\{andrea.nocentini, pierpaolo.panti\}@tecnolink.it}}} \\
    \vspace{0.2cm}
    {\large {\texttt{giaco.chiti@gmail.com}}}

\begin{minipage}{\textwidth}
\vspace{0.5cm}
\begin{titlepage}
\begin{abstract}
    \noindent
    We study the statistical properties of trainable agents moving in discrete space. After introducing the mathematical framework, we first analyze the dynamics of two completely random walkers, mutually competing in a chaser-target interaction scheme. The statistics of the encounters is analytically obtained and the predictions tested versus numerical simulations. We then move forward to extend the baseline case to agents capable of learning and adapting to an external reward signal, using reinforcement learning. Smart walkers morph the statistics of the encounter, to maximize their cumulated reward, as confirmed by combined numerical and analytical insights. More interestingly, configuration entropy proves a reliable proxy to gauge the acquired ability of the agents to cope with the assigned task when no other information about them (\textit{i.e.} reward signal, policy, etc) is present. We further test the proposed measure of learned skills by operating the Stockfish chess engine against a quasi-random untrained opponent. The obtained conclusions corroborate our claim. Summing up, our primary contribution is to propose and test a quantitative measure of agents’ awareness that naturally correlates with the inherent complexity of the task being performed.
\end{abstract}
\end{titlepage}
\vspace{0.5cm}
\end{minipage}
\end{center}
}]

\section{Introduction}

The study of random walkers has historically provided invaluable insights into a wide range of complex phenomena. Examples can be found in statistical physics \cite{NAGCHOWDHURY20221, PhysRevE.93.052111, PhysRevE.89.052147, PhysRevE.100.062310}, network theory \cite{Fortunatonet, CencettiNet, febbe2026Random, PhysRevE.103.042312}, robotics \cite{Tanrobotics} or finance \cite{RabiRW, Aydiner2019MoneyDI}. Random walkers often navigate in discrete spaces by using simple stochastic rules. Therefore, they offer a reference baseline to elaborate on a large plethora of complex dynamical schemes. These latter may account for adaptability, memory, or even strategic interactions to enable for a comprehensive description of the different phenomena under scrutiny. 

Particularly interesting is the setting where multiple walkers are made to simultaneously evolve in time, while populating the same spatial domain. As a relevant case study we will focus on the subset where just two walkers, a chaser and a target, are randomly relocating across the discrete space to which they belong and starting from the patch they are initially assigned to. The underlying space can be a regular lattice, extending in arbitrary dimensions, or a heterogeneous network: the same tools and formal attributes applies to either settings. For what it follows, and with no loss of generality, we choose to focus on a one-dimensional spatial domain, partitioned in discrete units, the cells that  the walkers can hit, ensuing successive randoms updates of their current state. This simple layout can be invoked as first descriptive scheme to model a large variety of physical systems, spanning from ecology to robotics, 
via gaming and multi-agents systems. More concretely, and to fix ideas, we shall hereafter make reference to the problem of trading, to which we took somehow inspiration. An order book is the list of orders, either manual or electronic, that a trading venue uses to 
to keep track of the interest of buyers and sellers in a specific financial item. In the crudest approximation possible, bids (buyers) and asks (sellers) place their respective orders, as positions on a one dimensional line where prices are ideally represented. Buyers populate the left portion of the line, while sellers are located on the right edge. Assuming the agents as mutually independent, and under the unlikely hypothesis of random updates of their open positions on the order book, the process can be brought back to the process of just two random walkers, wandering on a line. Clearly, this is an over-simplified scenario which does not account for strategic planning or intelligent interference among the involved actors. As we shall see, accommodating for some of those aspects define the actual scope of the present work. Before turning to elaborate on these important points, focus again on the aforementioned simplistic picture with just two walkers positioned on the opposite ends of an oriented line and moving as dictated by a time discrete Markov process. The quantities of interest are the (average) time of first meeting, given the initial conditions, and the position where the encounter eventually takes place. In the financial context, and disregarding the obvious limitations to which we alluded above and that  are intrinsic to the proposed schematic layout, the position where the walkers merge could flag for the price of the exchange. {\it Mutatis mutandis} the very same setting could apply to e.g. fencing, a sword fighting combat sport where opponents are facing each other and move back an forth on a pitch (which can be assimilated to a line), until a hit takes place (namely when one individual strikes the other). In this case, the time of collision is the relevant quantity, as each pool round lasts for a  finite amount.  Other contexts where the same conceptual framework virtually holds are pursuit-evasion games \cite{1067989}, predator-prey interactions \cite{Cooper}, cops-and-robbers scenarios \cite{BONATO20095588}, and even princess-monster games \cite{Alpern}. 

Further, it should be  remarked that the practical significance of understanding and optimizing meeting times is evident in multiple domains, such as stochastic surveillance strategies to promptly detect mobile intruders. Critical applications include environmental monitoring \cite{LIU2025121804}, emergency vehicle dispatching \cite{CHUNG2024101111}, traffic routing \cite{10546916}, and border patrol operations \cite{1656525}. Analyzing meeting times also contributes to a broader understanding of information flow in distributed systems \cite{SINGH2024111363}, self-stabilization mechanisms \cite{Zaiser}, and object similarity assessments in network contexts \cite{Jeh}. Early research on meeting times originated from self-stabilizing token management schemes \cite{Tetali}. In these schemes, exactly one processor in a distributed network possesses a token. This token grants it exclusive rights to change state or execute specific tasks. When two tokens meet, they merge into one. Israeli and Jalfon \cite{Israeli} proposed a randomized token-passing method. They obtained an exponential bound for the meeting time on general connected, undirected graphs. Coppersmith et al. \cite{Coppersmith} later refined this result by achieving a polynomial bound in terms of nodes count, leveraging pairwise hitting times between starting nodes and hidden vertices. In \cite{Coppersmith,Israeli,Tetali}, the adversarial token movement involves an adversary asynchronously moving only one of two tokens (e.g., pursuer-evader). The objective of the analysis is to maximize the meeting time. Bshouty et al. \cite{BSHOUTY1999259}, instead, provided bounds for the meeting time of multiple tokens, based on two-token meeting times. Finally, in George et al. \cite{george2018meeting}, a set of necessary and sufficient conditions was provided. These conditions characterize when the meeting time between a single pursuer and a single evader is finite for arbitrary Markov chains. In the same paper, the authors presented a closed-form solution to the meeting time by utilizing the Kronecker product of the transition matrices. 

Despite extensive research into these dynamics, most studies assume that all agents involved follow fixed, memory-less random-walk strategies. As anticipated above, this simplification overlooks scenarios where at least one agent can adapt strategically based on accumulated experience or environmental feedback. For the problem of book order forming in financial trading, it is clear that the updates are non random and follows a manifest tactical planning. Similarly, to make explicit reference to the second evocative example provided above, fencers do not relocate randomly but adjust their relative positions as reflecting the opponent's moves. To partially address this point and to take one step in the direction of accounting for strategic acting, we here consider the following scenario: one of the walkers implements an adaptive strategy, assimilated through reinforcement learning (RL)\cite{RL-An-Intro}. Specifically, we investigate qualitatively and quantitatively how the probability distribution of the first encounter changes (and the associated encounter times), as learning progresses. Reward is the crucial concept that shapes learning: the smart agents is given kudos (material or immaterial) in recompense for a somehow good performance, acting as positive reinforcement. For traders, higher rewards are gained when the encounter occur on the left most (resp. right most) portion of the lines, if the smart agent act as a buyer (resp. seller). As it could be guessed, non trivial change occurs when one walker transitions from a traditional stochastic policy to a learned, \textit{smart} policy. 

Assuming the above as a minimal though representative setting, we elaborate on a robust and quantitative approach to assess the level of awareness acquired by the agents' upon training. Specifically, we propose entropy as a viable measure of the agents' capacity as induced by experience. More precisely, entropy emerges as a quantitative tool to evaluate the degree of sophistication of the chosen policies and the complexity of the task under analysis. This is indeed the main result that we intend to deliver, the two agents setting providing the ideal framework to quantitatively elaborate in the sought direction. Notice that the above perspective contrasts with the Maximum Entropy framework commonly used in RL \cite{han2021max}, which is not here addressed. That framework promotes exploration and robustness by explicitly incorporating policy entropy into the reward function. Unlike the Maximum Entropy framework, where increased entropy encourages exploration of diverse actions and states for greater adaptability, our entropy measurement is computed ex post from the policy tensor itself. Therefore, it does not influence the smart agent training process.

Our contribution, thus, lies at the intersection of stochastic processes (Markov chains) and RL. We  characterize how adaptive strategies influence meeting-time distributions and  draw an intriguing connection with related entropy measures. To our knowledge, this integration of analytical meeting-time insights with RL-driven policy learning and entropy analysis constitutes a novel avenue of exploration. It enriches both theoretical understanding and practical implementations of adaptive pursuit.

The manuscript is structured as follows: In Sec. \ref{sec:model} we describe the model of two random walkers on a Markov chain as a game. In Sec. \ref{sec:dumbwalkers}, we present the simplest case of two random walkers. We  provide an analytical derivation of the probability distribution of the first encounter as a new result, and then we re-derive known results \cite{george2018meeting} through a novel and simpler proof detailed in Appendix \ref{app:proofs-of-closed-forms}. This result constitutes the benchmark reference for the subsequent analysis. In Sec. \ref{sec:smartwalkers}, we illustrate how to incorporate a smart strategy into a walker, outlining the reinforcement learning training procedure. Specifically, one of the two agents is adaptively trained to eventually develop an intelligent walk, under a reward that is being set a priori. At the end of the process, the smart walker has learned a strategy which is then applied during the testing phase: at each iteration the intelligent walker takes a move that is not dependent on past history. It only depends on its own position and on the position of the opponent, which keeps on performing an unbiased random walk. The pattern of possible moves is hence encoded in a generalized states space (the smart walker acts differently depending on its own position and on the position occupied by the opponent, at the time the move has to be taken). The dynamics is therefore exactly Markovian, namely it follows a stochastic process where the future state depends solely on the current state - in this case, of both walkers - not on the past history. This enables us to generalize the above formal analysis to the relevant setting where one (potentially two) walker(s) evolve following an intelligent strategic plan. The information-theoretic and thermodynamic metrics utilized in our analysis is also introduced. Sec. \ref{sec:numericalresult} presents our numerical results, achieved by training a random walker using linear, sinusoidal, and time-dependent reward policies. 
Moreover, in Sec. \ref{chess}, and motivated by the success of the above analysis, we speculate that configuration entropy can be employed as an effective measure of attitude or intelligent behaviour. To elaborate further in this direction we consider a real world application to table top games. Finally, we conclude in Sec. \ref{sec:discuss} by discussing the limitations and strengths of our research. The case of two simultaneously trained agents ( interacting via a predator- prey scheme) is shortly addressed in the annexed Appendix \ref{app:predator_prey} to show the versatility of the proposed approach, beyond the main setting here addressed.

\section{The general model}\label{sec:model}

\begin{figure*}[t]
    \centering
    \begin{tikzpicture}
        \foreach \i in {0,...,10}{
            \draw (\i,0) rectangle (\i+1,1);
            \node[below, blue] at (\i + 0.5, 0) {\the\numexpr\i-5\relax};
            \node[above, orange] at (10.5-\i, 1) {\the\numexpr\i-5\relax};
        }
        \draw[fill=orange] (1.5,0.5) circle (0.3);
        \draw[fill=blue] (9.5,0.5) circle (0.3);
        
    \end{tikzpicture}
    \caption{Illustration of the game environment for $N = 11$. Alice (orange) starts on the left, and Bob (blue) on the right. The scores displayed above and below the cells are drawn with the same color of the agent they refer to. Reflecting boundary conditions are applied at the edges. The depicted score distribution represents just one viable choice, among several other that we will set to explore all along the work.}
    \label{fig:environment}
\end{figure*}
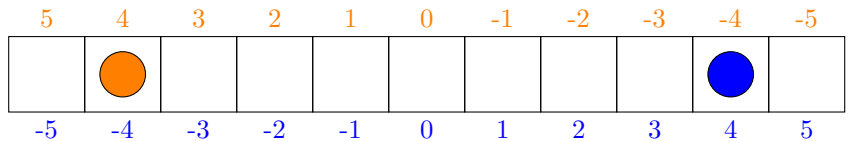

Consider a simple two-player game played on a one-dimensional grid with \(N\) discrete cells, numbered from left to right. Player one (Alice) is initially assigned to the left (as compared to the median point) domain, while player two (Bob) occupies a cell that pertains to the complementary portion of the available domain. The game unfolds in discrete turns, during which both players move simultaneously, according to specific rules. These latter can be random or reflect a lucid, thus seemingly intelligent, behaviour, as we will argue below. At each round, a player may move left,  right, or remain in place. The game ends when the two players come to occupy the same cell.

Each cell of the grid carries a payoff for both players, which is awarded only at the end of the game. More specifically, following an encounter, Alice and Bob receive an individual reward (established a priori, namely before the beginning of the game) and which reflects the specific location where the meeting took place. For the sake of clarity, let us refer to a particular payoff pattern: imagine Alice’s payoff decreases as the cell index increases (higher scores on the left), while Bob’s payoff increases with the cell index (higher scores on the right). If Alice is a smart being that seeks to maximize its own revenue, she will prevalently occupy the cells positioned in the leftmost portion of the grid, waiting for Bob to invade this latter domain. On the other hand, Bob is  incentivized to wonder among the cells that populate the rightmost domain, with a specular attitude as compared to that manifested by Alice. This  creates an obvious strategic tension:  cells that are most valuable to one player are the least valuable to the other. Since both players begin in regions favorable to their own payoff, the true challenge of the game is to pull the other into their own preferred side of the grid.

To ensure the walkers remain on the board, the edges of the grid (i.e., the leftmost and rightmost cells) are equipped with reflecting boundary conditions. If a player attempts to move beyond the board's limits, they keep still instead. Fig. \ref{fig:environment} provides 
a schematic layout of the illustrated setting. As a further ingredient, we assume that the two agents cannot compenetrate into each other. Stated different, they are prevented to mutually cross as follow a simultaneous update of their respective states. Hence, Alice will be  always found on the left of Bob. In the following we shall begin by analysing the setting where the two agents are assumed to execute unbiased random walks.
 
\section{Mathematical setting: two co-evolving random walkers}\label{sec:dumbwalkers}

\begin{figure*}[t]
    \centering
    \begin{subfigure}[b]{0.45\textwidth}
        \centering
        \includegraphics[width=\textwidth]{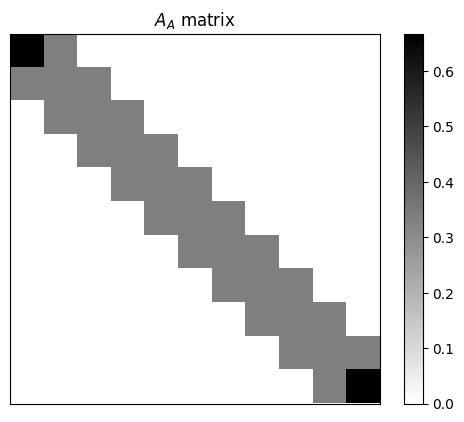}
        \caption{Alice's transition matrix.}
        \label{fig:sub1}
    \end{subfigure}
    \hfill
    \begin{subfigure}[b]{0.45\textwidth}
        \centering
        \includegraphics[width=\textwidth]{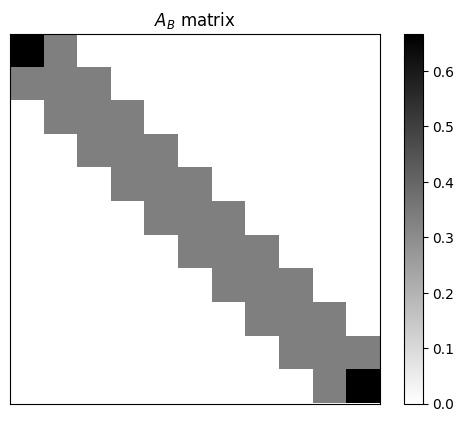}
        \caption{Bob's transition matrix.}
        \label{fig:sub2}
    \end{subfigure}
    \caption{Transition matrices for the two completely random agents. Reflecting boundary conditions are encoded within the matrices. At this stage, there is no interaction between the walkers, which hence execute two independent random walks.}
    \label{fig:matrices}
\end{figure*}

\begin{figure*}[t]
    \centering
    \includegraphics[width=0.7\textwidth]{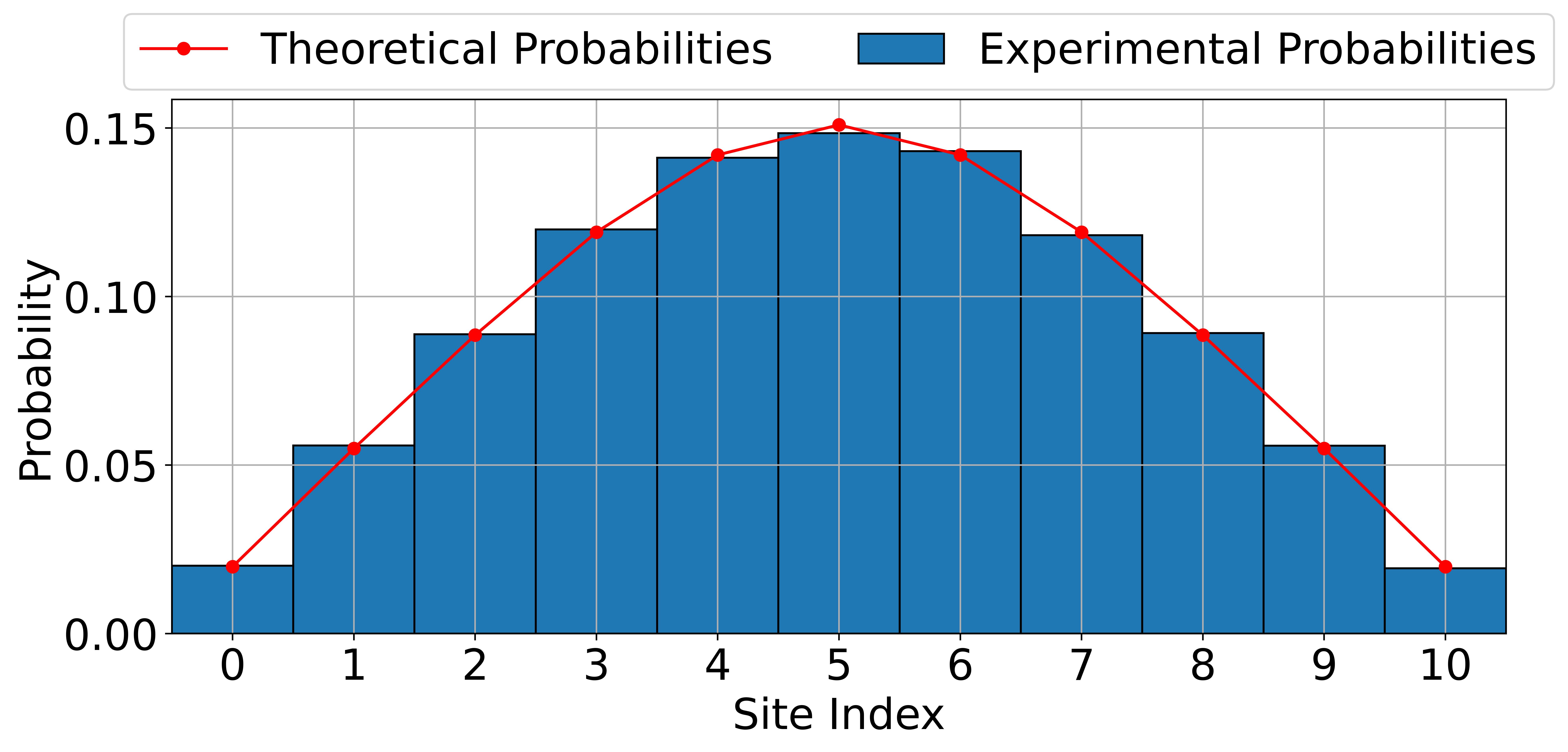}
    \caption{First encounter probability distribution for two random walkers under reflecting boundary conditions, with $N = 11$ and cell indices starting from $0$ (\(50{,}000\) simulated episodes). In the limit of infinitely many simulations, the empirical and theoretical distributions converge. Note that the tails of the distribution decrease approximately linearly, indicating a behavior that deviates from a quadratic profile.}
    \label{fig:Jacobi}
\end{figure*}

\begin{figure*}[t]
    \centering
    \begin{subfigure}[b]{0.45\textwidth}
        \centering
        \includegraphics[width=\textwidth]{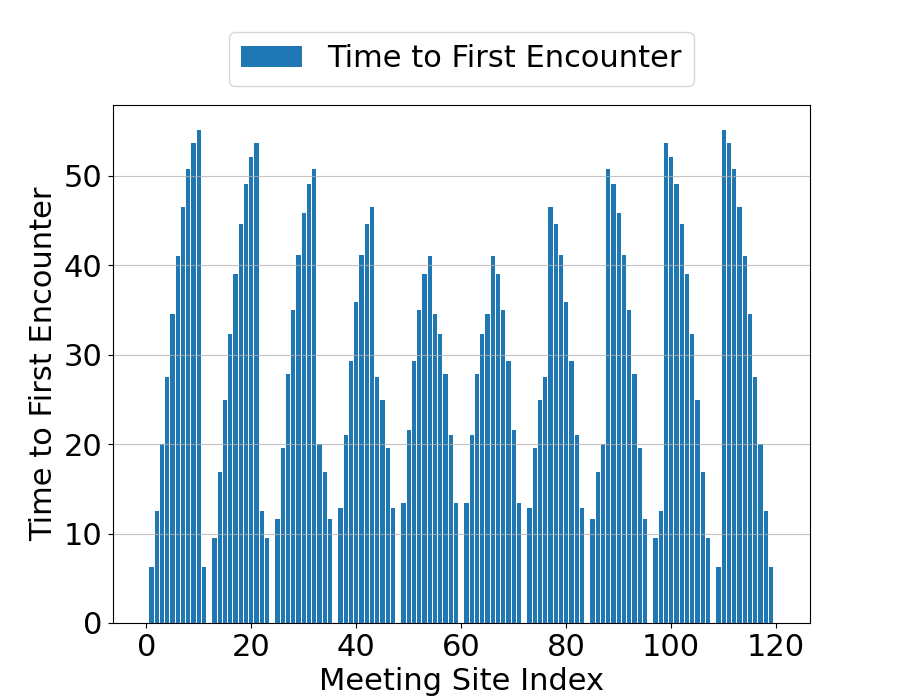}
        \caption{Bar plot of the $\bm{t}$ vector in \eqref{eq:t_vector_def}. The configuration index enumerates the elements of the vector in the tensor product space of possible $2$-walkers configurations.}
        \label{fig:random_first_enctimes_vector}
    \end{subfigure}
    \hfill
    \begin{subfigure}[b]{0.45\textwidth}
        \centering
        \includegraphics[width=\textwidth]{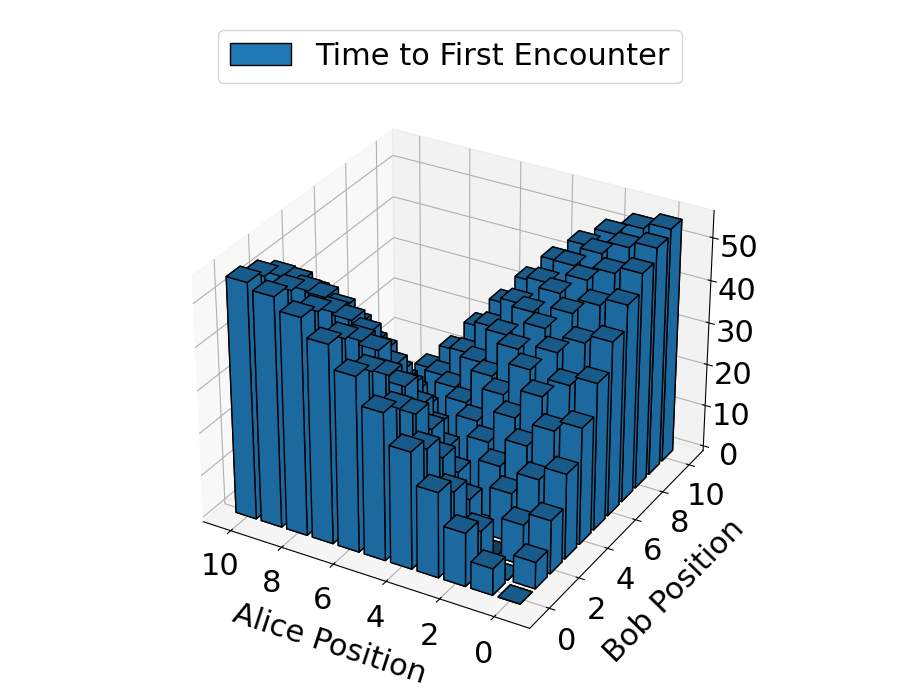}
        \caption{Bar plot of $\bm{t}$, factorized in the walkers' subspaces, using \eqref{eq:closed-form-t}.\\\\}
        \label{fig:random_first_enc_times_fact}
    \end{subfigure}
    \caption{Visualization of the vector $\bm{t}$ representing meeting times for two random walkers on a discrete space with $N = 11$. The left plot depicts directly the values of the elements of \(\bm{t}\).  The right plot instead shows a factorized view of $\bm{t}$ across the individual walker subspaces using Eq. \eqref{eq:closed-form-t}, resulting in a bar plot of mean meeting times (in game moves) as a function of the walkers' starting positions. The central trough indicates zero encounter times when both walkers start from the same cell.}
    \label{fig:random_first_enc_times_full}
\end{figure*}

In this section, we set to examine the case where the players behave non-strategically, which reduces the problem to two independent random walkers on a discrete lattice with reflective boundary conditions. Hence,  Alice (A) and Bob (B) do relocate with local jumps which occur with a uniform  probability: they can move right, remain still, or move left, each with an identical probability equal to \(1/3\). The only exception is at the leftmost and rightmost edges where the probability is split into 2/3 (to stay put) and 1/3 (to move towards the inside of the grid).

The motion of the walkers as specified above can be mathematically described in terms of the transition matrices $A_A$ and $A_B$, respectively associated to Alice and Bob. The entry $[A_k]_{ij}$ of matrix $A_k$, for $k \in \{A, B\}$, denotes the probability that walker $k$ transitions from cell $j$ to cell $i$ in a single step. By construction, only entries near to the diagonal are nonzero, reflecting the locality of movement. The structure of the ensuing sparse matrices is illustrated in Fig. \ref{fig:matrices}.  

Recall that the walkers evolve in time until they happen to simultaneously populate the same cell. At that point, the game ends and the corresponding reward (as reflecting the allocated payoff) is granted. To proceed further, it is useful to project the dynamics in an extended space where all possible combinations of 
Alice’s and Bob’s states are acknowledged, including those that materialize in an encounter. This is achieved by considering the tensor product of the individual transition matrices, denoted
\begin{equation}
    A = A_A \otimes A_B ,
\end{equation}
where $\otimes$ indicates the tensor product. 

Let $\bm{P}_A(t) \in \mathbb{R}^N$ and $\bm{P}_B(t) \in \mathbb{R}^N$ be the probability distributions over positions for Alice and Bob at time $t$, respectively. The full state of the two-walker system is then represented by the joint probability vector
\begin{equation}
    \bm{P}(t) = \bm{P}_A(t) \otimes \bm{P}_B(t) \in \mathbb{R}^{N^2},
\end{equation}
which evolves under the combined dynamics encoded in the matrix $A$.  

To explicitly account for the meeting conditions, the matrix $A$ as computed above need to be slightly modified. Recall that when the walker happens to share the same spatial location the game comes to an halt. In other words, $N$ states (those implying that Alice and Bob  have landed in the very same cell) of the $N \times N$ dimensional space should act as veritable absorbing traps. For each of such states, we erase the corresponding column and replace it with a zero uniform vector, but for the diagonal entry which is set to $1$.
This guarantees that once the walkers meet, in any of the $N$ cells where this can happen, they cannot evolve any further: Alice and Bob will be permanently frozen in the state they have reached, thus implying an effective cessation of the game.

Formally, let the joint state space be indexed by
\begin{equation}
    i = N \cdot x + y,
\end{equation}
where \(x\) and \(y\) denote the positions of Alice and Bob, respectively. This corresponds to the standard structure of tensor product indices (also known as Kronecker product indices, in this case). Here, we assume \(0 \leq x, y \leq N-1\). From this point onward, all indices will start from \(0\).

The set of meeting states is then defined by the condition $x = y$, i.e. the meeting states are indexed by the elements of the following \(\mathcal{M}\) set:
\begin{equation}
\mathcal{M} = \{ i = N \cdot k + k, \ k \in \{0,...,N-1\}\}  
\end{equation}

To enforce absorbing behavior at these states, we modify matrix $A$, as recalled above. Mathematically this amounts to cast the modified transition matrix $\tilde{A} \in \mathbb{R}^{N^2 \times N^2}$ as:
\begin{equation}
 \tilde{A}_{\cdot, i} =
\begin{cases}
\mathbf{e}_i & \text{if } i \in \mathcal{M}, \\\\
A_{\cdot, i} & \text{otherwise},
\end{cases}   
\end{equation}
with $A_{\cdot, i}$ being the $i$-th column of matrix $A$, and where $\mathbf{e}_i$ is the $i$-th column of the ${N^2 \times N^2}$ identity matrix $\mathbb{I} _{N^2 \times N^2}$. As already remarked, this construction guarantees that once the system reaches a meeting state, it remains there indefinitely.

To enforce the non-compenetration dynamics, we further modify $\tilde{A}$ by setting to zero the entries corresponding to forbidden switching processes, and by redistributing the probability mass equally among the remaining nonzero entries in each column of $\tilde{A}$. Notice that since we are focusing on identical walkers (i.e., walkers having the same probability distribution for their moves), this change has no effect on the first encounter probability distribution on the cells. On more general grounds, in the case of non-identical walkers, it would lead to a noticeable change in the first-encounter distribution.

The state of the system at a generic time \(t\) is described globally by the following dynamical equation:
\begin{equation}
\bm{P}(t) = \tilde{A}^{t-t_{in}} \bm{P}(t_{in}).
\label{eq:dynamics}
\end{equation}
$t_{in}$ denotes the initial time of the dynamics. Without loss of generality, we set $t_{in} = 0$.

By taking the limit as $t \to \infty$, we obtain the probability distribution of the first encounter between Alice and Bob at each cell of the board. We collect this information in a vector $\bm{\mathcal{P}} \in \mathbb{R}^N$, such that:

\begin{equation}
  \bm{\mathcal{P}}_k = \left[\lim _{t \to +\infty} \tilde{A}^{t} \bm{P}(0)\right]_{N\cdot k + k}
    \label{eq:trueevolution}   
\end{equation}

The shape of the resulting probability distribution $\bm{\mathcal{P}}$ is depicted in Fig. \ref{fig:Jacobi}), red solid line.  While it visually resembles a paraboloid , it is mathematically nontrivial. In fact, it is related to the Jacobi elliptic sine function \cite{stackpost}, as we shall illustrate below.

Given the formalism introduced above, the sought probability distribution for the first encounter can be computed without resorting on a limiting procedure. In fact, as demonstrated in Appendix \ref{app:proofs-of-closed-forms}, the following expression holds:

\begin{equation}
\bm{\mathcal{P}}_k = \sum _{\alpha=0}^{N^2-1}  M_{N\cdot k + k, \alpha}\big[\bm{P}(0)\big]_{\alpha}.
\label{eq:closed-form-p}
\end{equation}

where $\alpha$ is an index that spans the entire tensor product space. The matrix $M$, instead, denotes the inverse of the eigenvector matrix of $\tilde{A}$.

In Appendix \ref{app:proofs-of-closed-forms}, we also show how to derive a closed form for the average meeting time \(\tau_{a,b}\) for a configuration where Alice starts at position \(a\) and Bob at position \(b\). The resulting equation is:
\begin{equation}
\tau _{a,b} = [\bm{t}]_{N\cdot(b-1)+a},
\label{eq:closed-form-t}
\end{equation}
where $\bm{t}$ is a vector given by
\begin{equation}
\bm{t}=(\mathbb{I}-T)^{-1}\bm{c}.
\label{eq:t_vector_def}
\end{equation}
The matrix $T$ is defined as the transpose of $A_0$. The matrix $A_0$, in turn, is a modified version of \(A\), where the columns corresponding to the absorbing states have been replaced with zero vectors. The vector $\bm{c} \in \mathbb{R}^{N^2}$ is defined as the vector of all ones, except, again, at the entries corresponding to absorbing states, which are set to zero.
The vector of the average computed encounter times,  $\bm{t}$, is plotted in Fig. \ref{fig:random_first_enc_times_full}.

Notice that the results expressed in equations \eqref{eq:closed-form-p}, \eqref{eq:closed-form-t}, and \eqref{eq:t_vector_def}  remain also valid for walkers with non-homogeneous transition probabilities, as well as for walkers moving in higher dimensions, or on a graph. 

Starting from these premises, in the next Section we will set to consider the generalized case study where one walker is allowed to recast its own dynamics in light of the gained experience. The {\it smart agent} is exposed to different quenched rewards, as we shall highlight in the following, and training is implemented via reinforcement learning algorithms. For the sake of simplicity, we will primarily focus on contexts with just one trainable agent present. However, the introduced formalism can be straightforwardly extended to settings where multiple learners are present (as demonstrated in Appendix \ref{app:predator_prey}). We remind that this analysis is propaedeutic to challenging entropy as a reliable measure of the level of acquired agents' awareness.

\section{The case of a smart walker}\label{sec:smartwalkers}

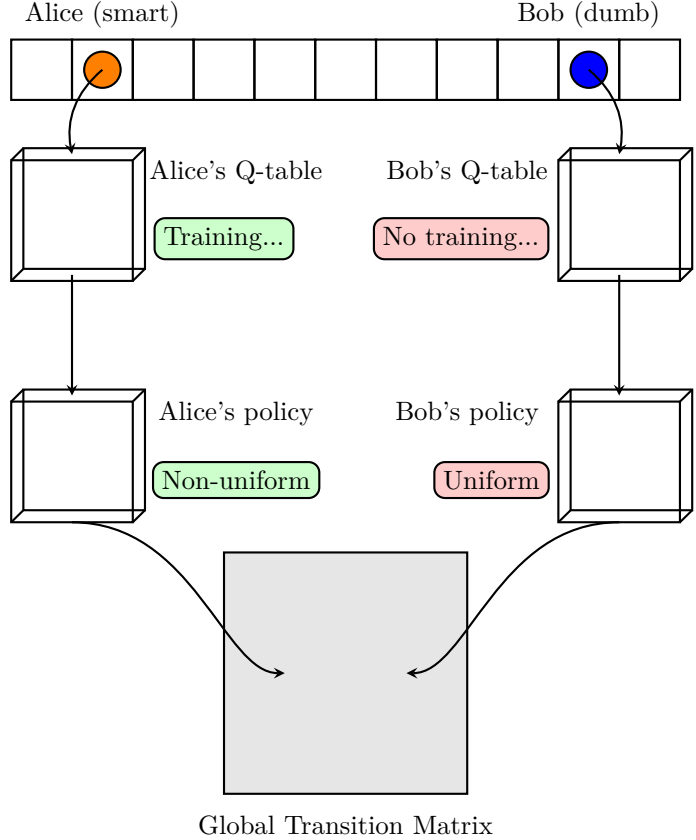
\begin{figure*}[t]
    \centering
    \begin{tikzpicture}[xscale=0.8, yscale=0.8, z={(0.1,0.1)},>=stealth,thick]

    \foreach \i in {0,...,10}{
        \draw (\i,0) rectangle (\i+1,1);
    }
    
    \draw[fill=orange] (1.5,0.5) circle (0.3) node[above=12pt] {Alice (smart)};
    \draw[fill=blue] (9.5,0.5) circle (0.3) node[above=12pt] {Bob (dumb)};

    \begin{scope}[canvas is xy plane at z=0]
        \coordinate (O1) at (0,-1);
        \coordinate (A1) at (2,-1);
        \coordinate (B1) at (2,-3);
        \coordinate (C1) at (0,-3);
    \end{scope}
    \coordinate (D1) at (0,-1,2);
    \coordinate (E1) at (2,-1,2);
    \coordinate (F1) at (2,-3,2);
    \coordinate (G1) at (0,-3,2);

    \begin{scope}[canvas is xy plane at z=0]
        \coordinate (O2) at (9,-1);
        \coordinate (A2) at (11,-1);
        \coordinate (B2) at (11,-3);
        \coordinate (C2) at (9,-3);
    \end{scope}
    \coordinate (D2) at (9,-1,2);
    \coordinate (E2) at (11,-1,2);
    \coordinate (F2) at (11,-3,2);
    \coordinate (G2) at (9,-3,2);

    \foreach \o/\a/\b/\c/\d/\e/\f/\g in {O1/A1/B1/C1/D1/E1/F1/G1, O2/A2/B2/C2/D2/E2/F2/G2}{
        \draw[thick] (\o) -- (\a) -- (\b) -- (\c) -- cycle;
        \draw[thick] (\d) -- (\e) -- (\f) -- (\g) -- cycle;
        \draw[thick] (\o) -- (\d);
        \draw[thick] (\a) -- (\e);
        \draw[thick] (\b) -- (\f);
        \draw[thick] (\c) -- (\g);
    }
    \node at (3.7,-1.2) {Alice's Q-table};
    \node at (7.5,-1.2) {Bob's Q-table};

    \draw[->,thick,bend right=30] (1.5,0.5) to (1,-0.9);  
    \draw[->,thick,bend left=30] (9.5,0.5) to (10,-0.9); 

    \node[fill=green!20,draw,rounded corners] at (3.5,-2.3) {Training...};
    \node[fill=red!20,draw,rounded corners] at (7.4,-2.3) {No training...};

    \begin{scope}[canvas is xy plane at z=0]
        \coordinate (O3) at (0,-5);
        \coordinate (A3) at (2,-5);
        \coordinate (B3) at (2,-7);
        \coordinate (C3) at (0,-7);
    \end{scope}
    \coordinate (D3) at (0,-5,2);
    \coordinate (E3) at (2,-5,2);
    \coordinate (F3) at (2,-7,2);
    \coordinate (G3) at (0,-7,2);

    \begin{scope}[canvas is xy plane at z=0]
        \coordinate (O4) at (9,-5);
        \coordinate (A4) at (11,-5);
        \coordinate (B4) at (11,-7);
        \coordinate (C4) at (9,-7);
    \end{scope}
    \coordinate (D4) at (9,-5,2);
    \coordinate (E4) at (11,-5,2);
    \coordinate (F4) at (11,-7,2);
    \coordinate (G4) at (9,-7,2);

    \foreach \o/\a/\b/\c/\d/\e/\f/\g in {O3/A3/B3/C3/D3/E3/F3/G3, O4/A4/B4/C4/D4/E4/F4/G4}{
        \draw[thick] (\o) -- (\a) -- (\b) -- (\c) -- cycle;
        \draw[thick] (\d) -- (\e) -- (\f) -- (\g) -- cycle;
        \draw[thick] (\o) -- (\d);
        \draw[thick] (\a) -- (\e);
        \draw[thick] (\b) -- (\f);
        \draw[thick] (\c) -- (\g);
    }
    \node at (3.7,-5.2) {Alice's policy};
    \node at (7.5,-5.2) {Bob's policy};
    
    \node[fill=green!20,draw,rounded corners] at (3.7,-6.3) {Non-uniform};
    \node[fill=red!20,draw,rounded corners] at (7.9,-6.3) {Uniform};

    \draw[->,thick] (1,-2.9) -- (1,-4.9);
    \draw[->,thick] (10,-2.9) -- (10,-4.9);

   \draw[fill=gray!20,thick] (3.5,-11.5) rectangle (7.5,-7.5) node[pos=.5, below=50pt]{Global Transition Matrix};
   \draw[->,thick] (1,-7) .. controls (3,-7) and (3.5,-9.5) .. (4.5,-9.5);
   \draw[->,thick] (10,-7) .. controls (8,-7) and (7.5,-9.5) .. (6.5,-9.5);

\end{tikzpicture}
    \caption{Generalized game environment for smart walkers. Each agent is equipped with a Q-table, which can be mapped to a corresponding policy tensor via Eq. \eqref{eq:temperature_softmax}. Initially, both policies correspond to a random walk. Over repeated games, Alice updates her Q-table using Eq. \eqref{eq:q-learning}, thereby modifying her policy tensor and inducing a new dynamic. Both agents’ policies can be used to construct the global transition matrix $A$, as detailed in Appendix \ref{app:get_global_A_from_policy}.}
    \label{fig:smart_walkers}
\end{figure*}

\begin{figure*}[t]
    \centering
    \begin{subfigure}{0.32\textwidth}
        \centering
        \includegraphics[width=\linewidth]{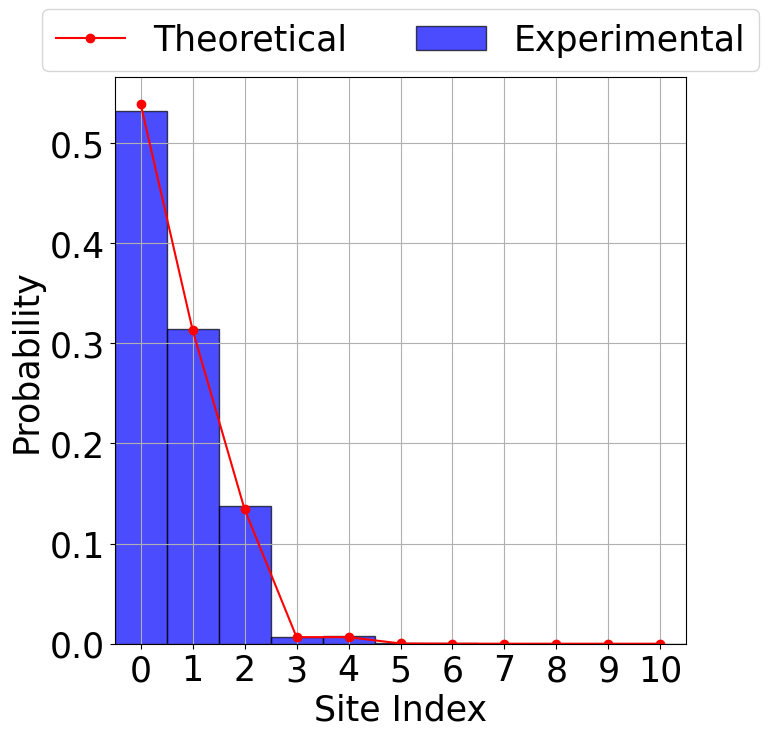}
        \caption{First encounter probability distribution after  Alice under linear reward signal \eqref{eq:linear_reward}.}
        \label{fig:tpd1}
    \end{subfigure}
    \hfill
    \begin{subfigure}{0.32\textwidth}
        \centering
        \includegraphics[width=\linewidth]{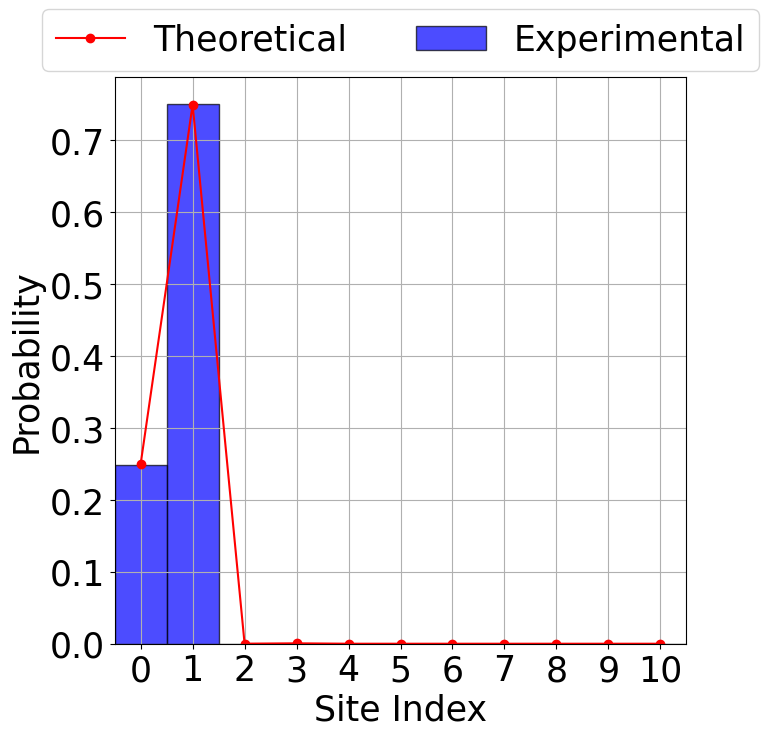}
        \caption{First encounter probability distribution after training Alice under linear time-dependent reward \eqref{eq:time_dependent_linear_reward}.}
        \label{fig:tpd2}
    \end{subfigure}
    \hfill
    \begin{subfigure}{0.32\textwidth}
        \centering
        \includegraphics[width=\linewidth]{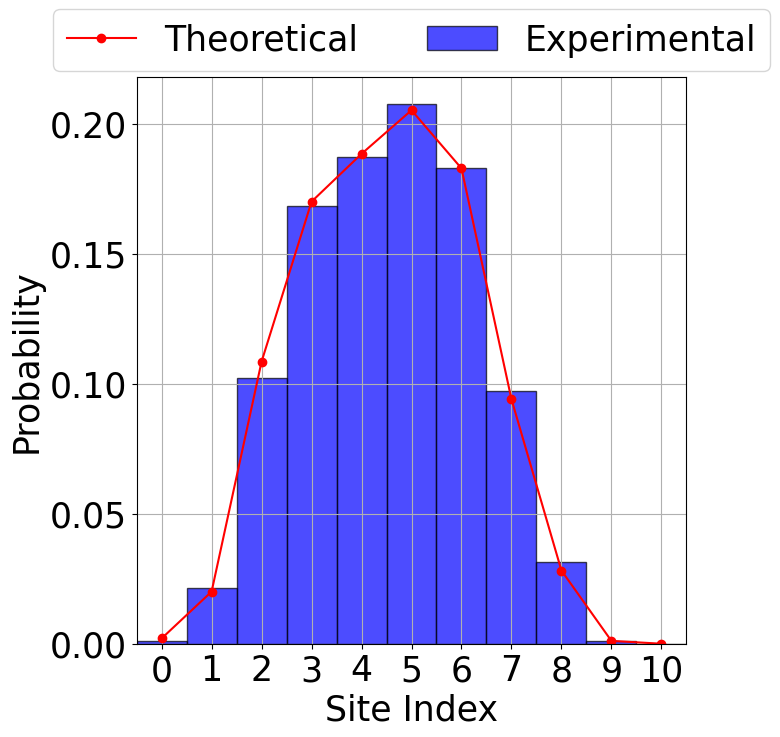}
        \caption{First encounter probability distribution after training Alice under sinusoidal reward signal \eqref{eq:sinusoidal_reward}.}
        \label{fig:tpd3}
    \end{subfigure}
    \hfill
    \caption{First encounter probability distributions, derived both numerically (\(10{,}000\) simulated games after learning) and analytically with \eqref{eq:closed-form-p}. Each bin of the bar plots corresponds to a different site in the game space. Stated differently, we do not impose any a posteriori binning; the level of displayed coarse graining reflects the specificity of the examined problem. The red curves superimposed to the blue bar plots refers to the theoretical prediction \eqref{eq:closed-form-p}, while the bar plot reports on the simulation’s result. The theoretical curve  takes values on the very same discrete support that defines the available configuration space. No fitting parameters are adjusted when superposing the aforementioned theoretical and numerical curves.}
    \label{fig:trained_pd}
\end{figure*}

\begin{figure*}[t]
    \centering
    \begin{subfigure}{0.32\textwidth}
        \centering
        \includegraphics[width=\linewidth]{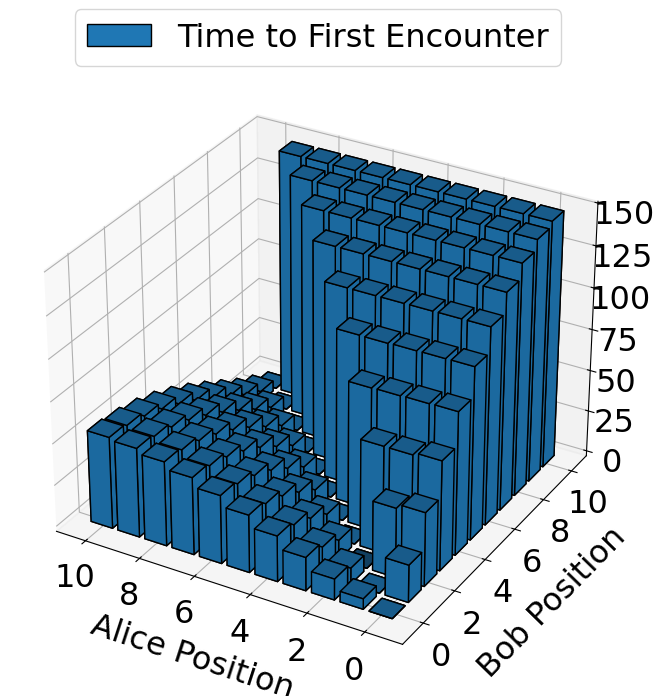}
        \caption{Mean first encounter times after training Alice under the linear reward \eqref{eq:linear_reward}.}
        \label{fig:ttimes1}
    \end{subfigure}
    \hfill
    \begin{subfigure}{0.32\textwidth}
        \centering
        \includegraphics[width=\linewidth]{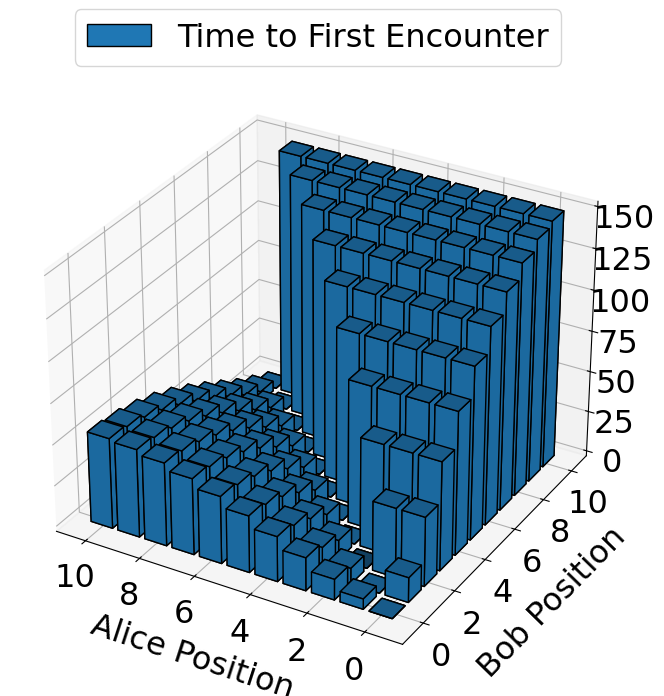}
        \caption{Mean first encounter times after training Alice under the time-dependent reward \eqref{eq:time_dependent_linear_reward}.}
        \label{fig:ttimes2}
    \end{subfigure}
    \hfill
    \begin{subfigure}{0.32\textwidth}
        \centering
        \includegraphics[width=\linewidth]{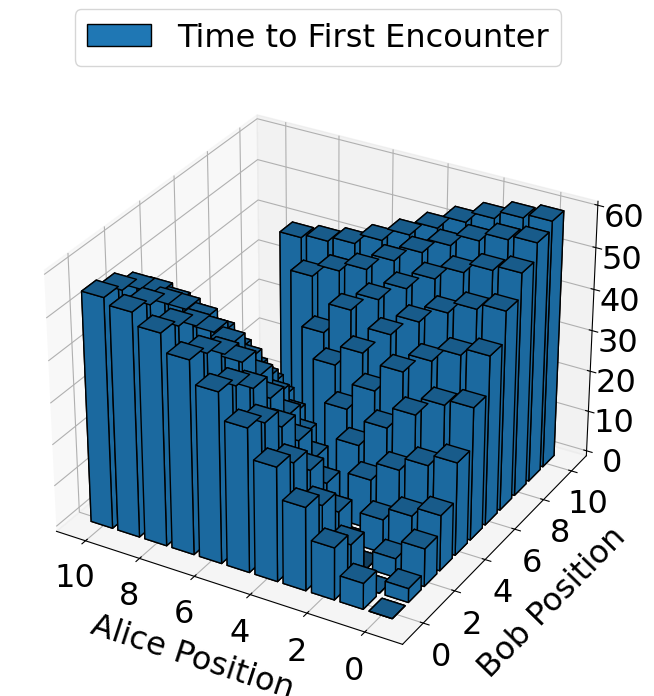}
        \caption{Mean first encounter times after training Alice under the low frequency sinusoidal reward \eqref{eq:sinusoidal_reward}.}
        \label{fig:ttimes3}
    \end{subfigure}
    \hfill
    \caption{{Mean first encounter times, derived analytically with \eqref{eq:closed-form-t}.}}
    \label{fig:trained_times}
\end{figure*}

\begin{figure*}[t]
    \centering
    \begin{subfigure}{0.32\textwidth}
        \centering
        \includegraphics[width=\linewidth]{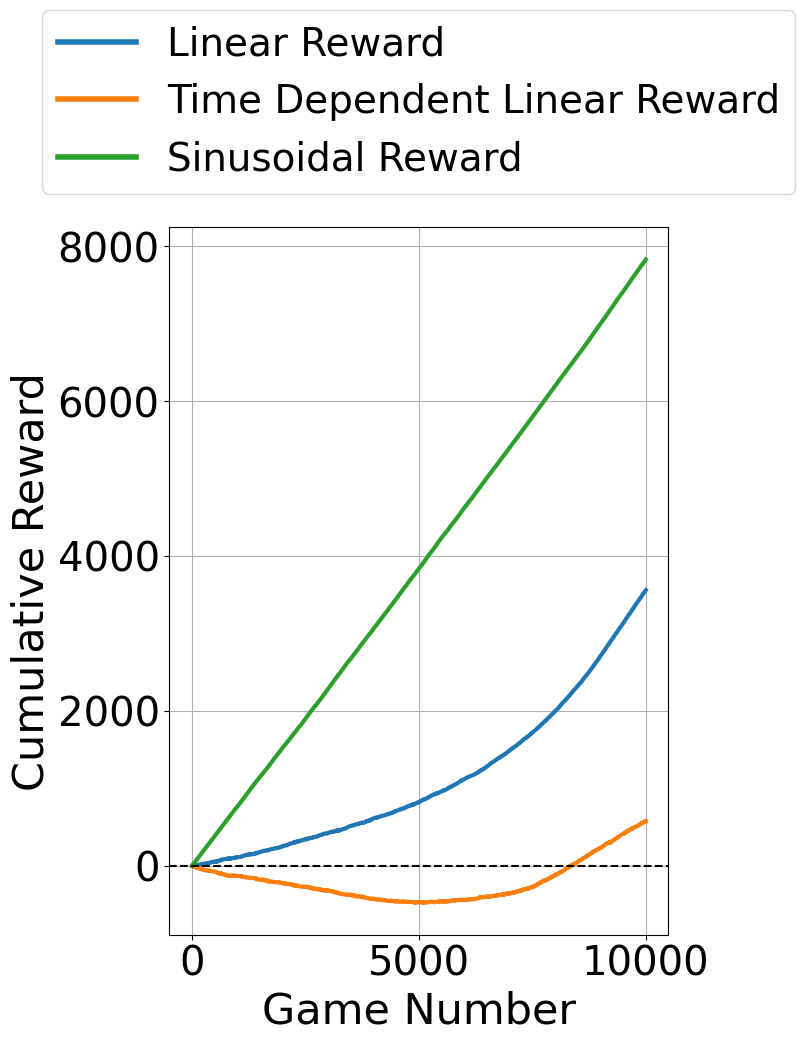}
        \caption{}
        \label{fig:cumulative_rewards}
    \end{subfigure}
    \hfill
    \begin{subfigure}{0.32\textwidth}
        \centering
        \includegraphics[width=\linewidth]{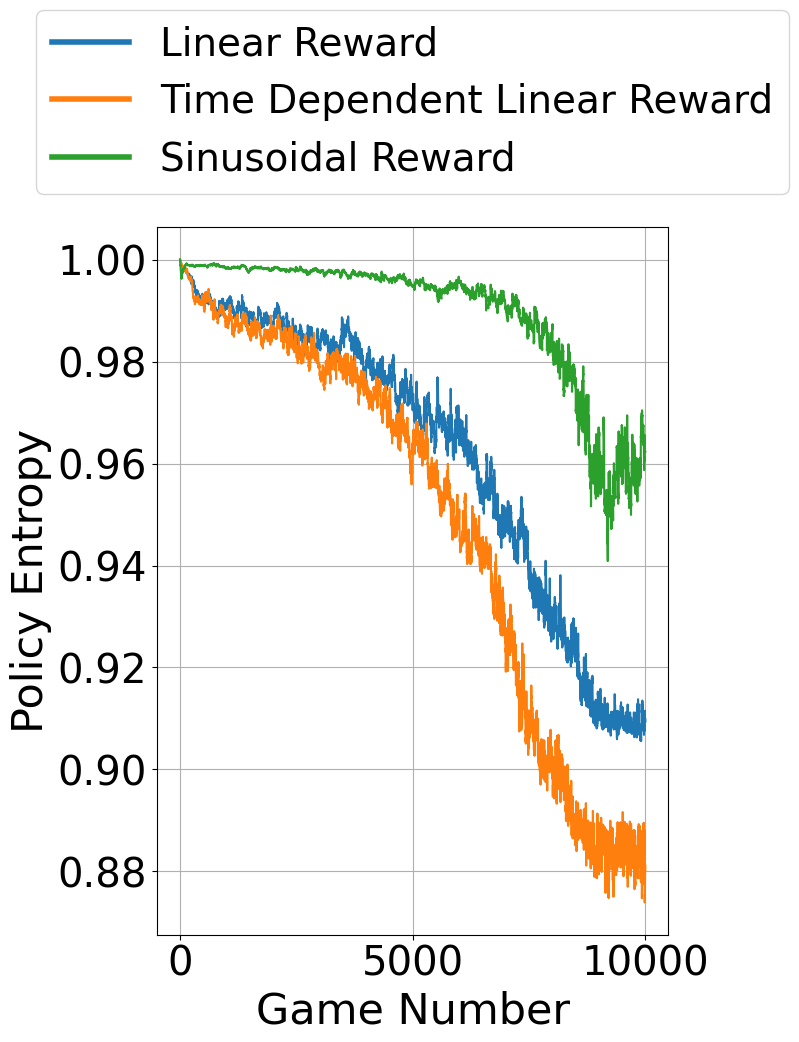}
        \caption{}
        \label{fig:policy_entropies}
    \end{subfigure}
    \hfill
    \begin{subfigure}{0.32\textwidth}
        \centering
        \includegraphics[width=\linewidth]{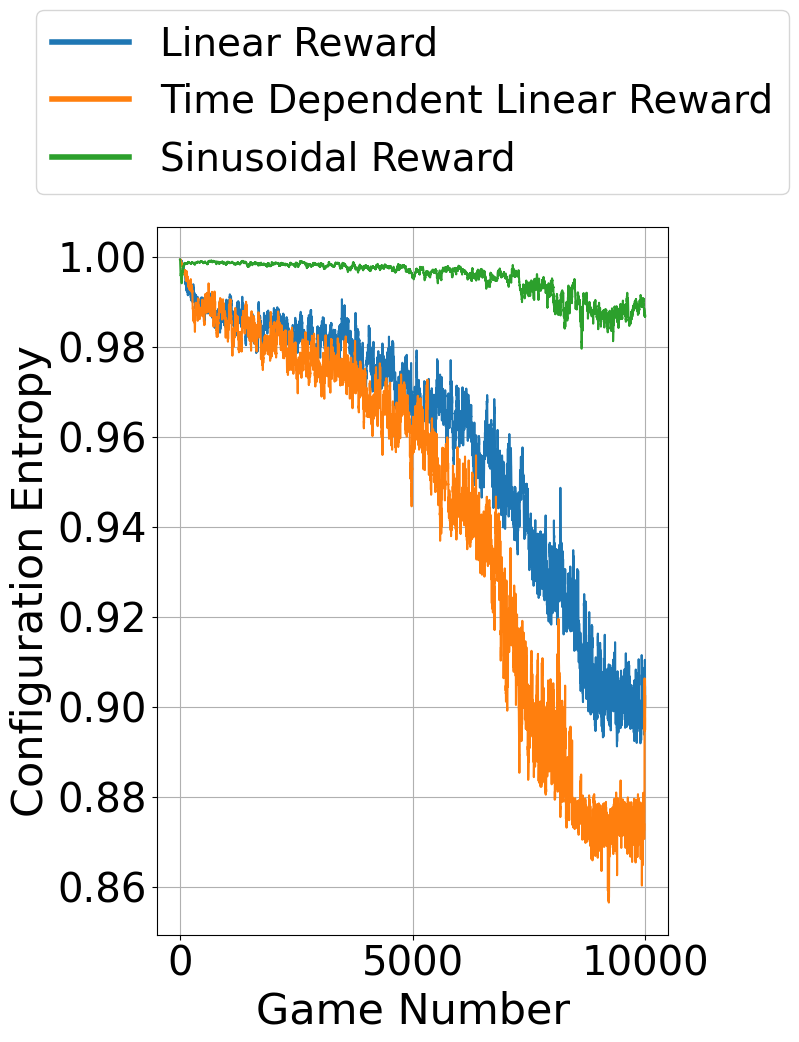}
        \caption{}
        \label{fig:configuration_entropies}
    \end{subfigure}
    \hfill
    \caption{{(a) Cumulative rewards obtained by Alice during training. The strongest performance belongs to the walker trained on the sinusoidal reward \eqref{eq:sinusoidal_reward}, while the worst result is associated with the time-dependent linear reward. (b) Normalized values of the cumulative Shannon policy entropy of Alice, as a function of the game number, for different reward functions. All the entropy profiles start at the maximum possible value, corresponding to Alice having learned nothing yet (and so performing a random walk). As the games proceed, the entropy value drops, thus signaling the encoding of information in Alice's policy, as resulting from the training. (c) Normalized values of the configuration entropy. This correlates strongly with Alice's Shannon policy entropies. Also, the relative ordering between the curves pertaining to different reward functions is maintained in all three reported measures (cumulative reward, policy entropy, configuration entropy).}}
\end{figure*}

In the following we shall consider the setting where one agent gets trained to cope with the assigned task (chasing the target), with a suitably tailored reward that breaks spatial symmetry. We will begin our report on this generalized framework by illustrating the adopted training scheme. 

\subsection{The training pipeline}

Before analyzing the behavior of adaptive agents, we first introduce the employed learning mechanism. This latter allows a walker to develop a smart policy. Unlike the purely random behavior considered earlier, a smart agent uses reinforcement learning (RL) to improve its decision-making over time by maximizing the expected reward. Through repeated interactions with the environment, the agent adjusts its strategy based on past experience. It, thus, learns to balance exploration and exploitation in pursuit of more favorable outcomes.

To this end, we associate with each agent a new type of tensor commonly referred to as a \textit{Q-table}, following the terminology of reinforcement learning. In our setting, the Q-table is represented as a three-dimensional tensor $Q \in \mathbb{R}^{N \times N \times 3}$. The first two dimensions correspond to the full joint state space of the system.  Each state is denoted by the pair $(i, j)$, representing the positions of Alice and Bob, respectively. The third dimension indexes the agent's available actions (i.e., moving left, staying in place, or moving right).

The Q-table allows an agent to learn and exhibit intelligent behavior by conditioning its actions not only on its own position but also on the position of the other player. This represents a key difference from the earlier transition matrices $A_A$ and $A_B$, which encode only self-position-dependent stochastic behavior.

With this framework in place, our objective is to apply standard reinforcement learning techniques to train only one player (Alice), allowing her to learn a policy that maximizes long-term reward. Bob, in contrast, behaves as a standard random walker governed by uniform stochastic movement. This simple setup allows us to study the learning process of smart walkers in a simple fixed environment, as well as the deviations of their dynamics from the baseline - as discussed in the preceding Section - due to learning.

To train Alice’s Q-table, we employ the standard Q-learning algorithm \cite{RL-An-Intro}, described as:
\begin{equation}
    q(s,a) \leftarrow (1-\alpha)q(s,a) + \alpha\big[r + \gamma \max _{a'} q(s',a')\big],
    \label{eq:q-learning}
\end{equation}
where the state–action quality tensor (Q-tensor) is updated using \textit{q-learning}: a temporal-difference learning rule, with a one-step look-ahead into the future. The parameter $\alpha \in [0,1]$, known as the learning rate, determines the extent to which new information overrides prior knowledge. It governs how aggressively the Q-values are updated based on new experiences. The variable $r$ denotes the reward received by the agent after taking action $a$ in state $s$; in our setup, this corresponds to the score assigned at the resulting meeting position. If no score is gained, the reward is simply set to zero. Finally, the parameter $\gamma \in [0,1]$, known as the discount factor, controls the agent’s preference for future rewards over immediate ones. A value of $\gamma$ close to 1 places more emphasis on long-term returns, while values closer to 0 make the agent more short-sighted; this can be also leveraged to promote general rapidity (for example quick encounters in chaser-target games). The symbols $s'$ and $a'$ represent the subsequent state and actions following the transition from $s$ via action $a$. We set $\gamma = 1$ and $\alpha = \frac{1}{2}$ in our simulations.

We define the agent’s policy $\pi(a|s)$ as a probability distribution over available actions given a state. This latter distribution is obtained by means of the so called \textit{Boltzmann exploration paradigm} \cite{amin2021}, which applies a weighted softmax function $S_\beta$ to the Q-values of the given state. More concretely, we stipulate:
\begin{equation}
    \pi(a|s) = S_{\beta}(\bm{q}(s,\bm{a}))_a
\end{equation}
where $\bm{q}(s,\bm{a})$ stands for the vector of all the qualities corresponding to all the actions $\bm{a}$ that the model can take in state $s$. The weighted softmax $S_\beta$ is defined as
\begin{equation}\label{eq:temperature_softmax}
S_\beta(\mathbf{x})_i = \frac{e^{\beta x_i}}{\sum_{j=1}^{n} e^{\beta x_j}},   
\end{equation}
where $\beta > 0$ is the inverse temperature parameter controlling the trade-off between exploration and exploitation. Larger values of $\beta$ place more weight on actions with higher Q-values, promoting exploitation, while smaller values yield a more uniform distribution, encouraging exploration. This formulation is mathematically equivalent to the Boltzmann distribution, hence the name.

As it is customary during training, we apply a linear temperature annealing schedule, where the inverse temperature is progressively increased as the training progresses. Formally, at episode $e$ the inverse temperature is given by
\begin{equation}
    \beta(e) = \frac{1}{\max\left(\frac{1}{\beta_+}, \; \frac{1}{\beta_0}\left(1 - \frac{e}{E}\right)\right)},
\end{equation}
where $\beta_0$ denotes the initial inverse temperature, $\beta_+$ is the upper threshold, and $E$ is the total number of episodes.

From the agents' policy tensors, it is possible to construct the global transition matrix of the Markov process (see Appendix \ref{app:get_global_A_from_policy}). Figure \ref{fig:smart_walkers} provides a schematic overview of the smart walker framework. This matrix can then be used in conjunction with equation \eqref{eq:closed-form-p} and \eqref{eq:closed-form-t} to compute all physical observables of interest. 

In each simulation presented in the next section, we compute a set of physically meaningful observables. First, we examine the first encounter probability distribution over the $N$ cells. This quantity, obtained via equation \eqref{eq:closed-form-p}, describes where meetings are most likely to occur. Next, we consider the distribution of meeting times across all $N^2$ possible starting configurations. This is computed using equation \eqref{eq:closed-form-t}.  We also compute the normalized thermodynamical entropy of the global Markov process, defined over the stationary distribution $\bm{\phi}_1$ (the eigenvector corresponding to eigenvalue \(1\)) of the transition matrix $A$ (not the $\tilde{A}$, but still forbidding the compenetration processes):

\begin{equation}
S_T(\bm{\phi}_1) = - \frac{1}{|\mathcal{C}|}\sum_{c \in \mathcal{C}} \phi_c \log \phi_c,
\label{eq:configuration_entropy}
\end{equation}
{where $\mathcal{C}$ stands for the set of all possible configurations for the two agents}.

At last, to evaluate the information content of the learned policies, we will be interested in the normalized cumulative Shannon entropy of the trained policy tensor:

\begin{equation}
S_S(\pi) = -\frac{1}{|\mathcal{S}|\log|\mathcal{A}|}\sum_{s \in \mathcal{S}} \sum_{a \in \mathcal{A}} \pi (a|s)\log \pi(a|s),
\label{eq:policy_entropy}
\end{equation}
{where $\mathcal{S},\mathcal{A}$ represent the set of possible states and the set of possible actions, respectively. Notice that even if $\mathcal{C}$ and $\mathcal{S}$ originate from different frameworks - thermodynamical versus game-theoretical - they are \textit{de facto} the same set.}

In the following section, we compare these quantities across different learning environments. As already mentioned, only Alice is allowed to learn, while Bob remains a simple, unintelligent, random walker.

\subsection{The imprint of learning on the recorded statistics}\label{sec:numericalresult}

In this section, we present our numerical analysis, where Alice was trained under three distinct reward signals:
\begin{itemize}

\item \textbf{Linear reward:} decreasing with the meeting site index \(n \in \{0, \dots, N-1\}\):
\begin{equation}
    r(n) = 1 - \frac{2n}{N-1}
    \label{eq:linear_reward}
\end{equation}

\item \textbf{Time-dependent linear reward:} initially identical to the linear case, but decaying further with an additive penalty proportional to the game move index \(t\):
\begin{equation}
    r(n) = 1 - \frac{2n}{N-1} - \lambda t , \quad \lambda \in \mathbb{R}^+
    \label{eq:time_dependent_linear_reward}
\end{equation}

\item \textbf{Sinusoidal reward:} a low-frequency signal designed to approximate the natural first-encounter probability distribution:
\begin{equation}
    r(n) = \sin \left( \frac{\pi n}{N - 1} \right)
    \label{eq:sinusoidal_reward}
\end{equation}

\end{itemize}
The  reward profile \eqref{eq:sinusoidal_reward} was chosen as it resembles the natural encounter profile of completely random, untrained, agents. The other two linear profiles were instead selected to sketch a toy zero-sum scenario, with or without time pressure. One could interpret, for example, the two walkers as a buyer-seller couple, moving in price space during trading, as already mentioned in the introductory section.
The agent (Alice) was trained for $10{,}000$ episodes under each reward function, using an inverse temperature parameter $\beta$ that started at an initial value of $1$ and was capped at an upper threshold of $10$. This training regime yielded dynamical systems of the chaser-target typology with substantially morphed first-encounter probability distributions (see Figure \ref{fig:trained_pd}) and mean first-encounter times (see Figure \ref{fig:trained_times}). In the aforementioned Figures, the histogram refer to actual simulation of the learned stochastic dynamics. The solid (red) lines stand for the analytical solutions, computed following the procedure introduced in Section \ref{sec:dumbwalkers}, and by making use of the policy table to which Alice converged upon learning, subject to the different assumed rewards. A comment is mandatory at this point. As also anticipated in the introduction, the problem of a smart walker chasing (post training) a target opponent (which still performs pure random walk) is intrinsically Markovian, when examined in the expanded space of the $N^2$ states  for the co-evolving pair.  No approximations are involved in the formal analysis reported above, which is therefore exact.

In e.g. panel (a) of Figure \ref{fig:trained_pd} a linear reward signal decreasing with the lattice site is assumed: clearly, the training biases the encounter towards the leftmost sites of the collection, where the reward for the smart walker is largest. This comes with no surprise (and similar considerations apply to the results displayed in panel (b) and (c) of Figure \ref{fig:trained_pd}) and for this reason we do not indulge in extensive comments related to the obtained optimal solutions. The interesting fact is that we can provide solid analytical arguments to benchmark the numerical experiments: this is the message delivered by our preliminary tests and which sets the ground for the successive analysis.

The dynamics during training reveal additional insights, as we move in the direction to comparatively evaluate the gained agents' ability in light of the complexity of the supplied tasks. Figure \ref{fig:cumulative_rewards} shows the cumulative rewards as a function of the training episode, while Figure \ref{fig:policy_entropies} depicts the Shannon entropy of the policy tensors, also tracked across episodes.

As evident from the Shannon entropy values in Figure \ref{fig:policy_entropies}, the time-dependent reward induces the highest information encoding in the agent's policy tensor. In contrast, the sinusoidal reward results in the lowest information content. This observation aligns with the other available data: the sinusoidal game is evidently easier, as indicated by the fact that the reward structure closely resembles the behavior of an unintelligent agent, mirroring the random encounter probability distribution. Conversely, the time-dependent game is the most challenging, and requires substantial learning to be mastered.

From the preceding analysis we conclude that even in a simple one-dimensional setting higher performances do not necessarily imply an higher amount of information learned. To gauge by how much an agent differs from a random player access to the policy information entropy is crucial. {In this sense policy information entropy gives a measure of the "\textit{intelligence}" acquired by the agent during training.}

In many real-world contexts, however, the policy tensor is not explicitly defined or directly accessible. For instance, biological agents such as bacteria in a Petri dish do not expose a neatly structured policy from which Shannon entropies can be computed. Remarkably, the configuration entropy can instead be estimated by simply observing the agent’s dynamics in the environment. This observation is particularly valuable, as the configuration entropy of trained walkers appears to correlate strongly with their policy entropy under consistent reward signals. This relationship is evident when comparing the Shannon policy entropies reported in Figure \ref{fig:policy_entropies} with the configuration entropies defined in \eqref{eq:configuration_entropy} and shown in Figure \ref{fig:configuration_entropies}.

\section{Configuration Entropy as a measure of skill}\label{chess}

\begin{figure*}[t]
    \centering
    \includegraphics[width=0.6\linewidth]{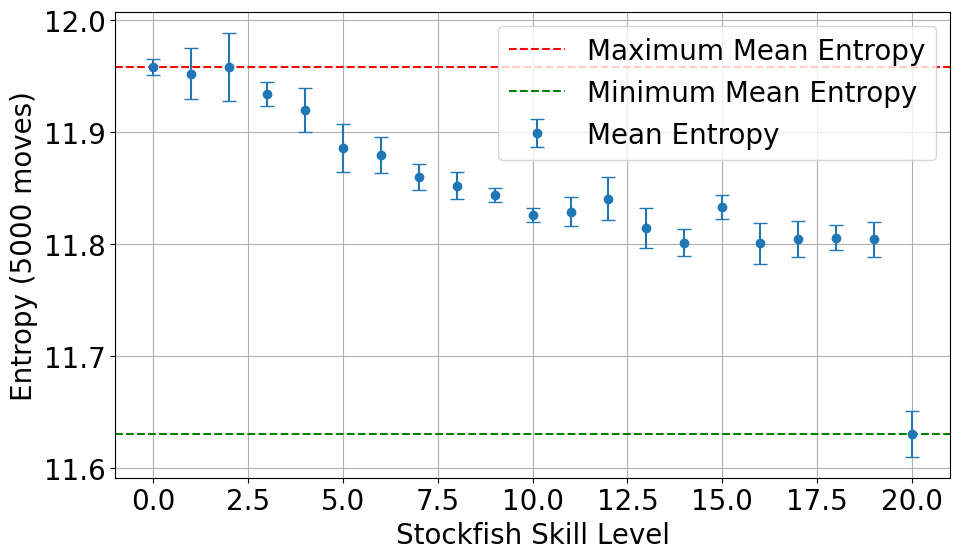}
    \caption{{In this figure the non-normalized configuration entropy values are reported as a function of Stockfish skill level. Each point is calculated by observing Stockfish play against a quasi-random opponent for $5,000$ moves (restarting the game, upon game conclusion, if necessary). This procedure was repeated $3$ times to extract the mean and its standard deviation. Note that entropy values appear to monotonically descend with Stockfish skill level, suggesting that they could be used a measure of inherent ability of the smart agents. Note also the entropy value gap between level $19$ and $20$: its presence is not an artifact as explained in the main body of the text.}}
    \label{fig:stockfish_entropy}
\end{figure*}

Building on the results presented in the previous section, it is thus tempting to interpret configuration entropy as an indirect proxy of the \textit{"intelligence"} of an agent. Notice that 
the above interpretation is solely justified for the context here analyzed, where just one agent (in a discrete state space) can learn from experience when facing a stereotyped (or unintelligent) opponent. To further challenge this working hypothesis, we move beyond the simple walkers framework to consider a richer class of environments, more complex and widely studied. These are tabletop games.

Among these, we select chess as a case study. Chess provides a well-defined, high-dimensional decision-making environment that has been extensively analyzed both in artificial intelligence and cognitive science. Moreover, it offers a practical advantage: the existence of a state-of-the-art synthetic player, the Stockfish engine. Stockfish is one of the strongest chess engines available, consistently outperforming even the best human players. Crucially for our purposes, Stockfish can also be configured to operate at different skill levels, enabling controlled experiments across a spectrum of agent competencies.

Specifically, Stockfish provides a range of skill levels indexed from 
\(0\) to \(20\). At level \(20\), the engine runs at its full strength, representing the unaltered policy of the underlying system. Levels \(0\) through \(19\), by contrast, correspond to progressively weakened versions of the engine, deliberately handicapped in order to approximate play at lower skill levels and thereby allow for more meaningful interaction with human players. This tunable degradation of performance makes Stockfish an ideal testbed for exploring the relationship between an agent's intelligence and its configuration entropy.

Figure \ref{fig:stockfish_entropy} reports the results of our experiments across all Stockfish skill levels. For each level, we simulated \(5{,}000\) moves against a quasi-random, unskilled opponent, in direct analogy with the procedure adopted in the previous section. During these simulations, the {board was started at the conventional initial state and a game was played until one of the opponents won or the maximum number of moves was reached.} Games were recorded, and the probability of encountering a particular board configuration, denoted by \(p_c\), was estimated as the ratio between the number of times that configuration occurred and the total number of configurations observed.  

From these empirical distributions, we computed the configuration entropy according to equation \ref{eq:configuration_entropy}. In this analysis, we chose to omit the normalization factor \(1/|\mathcal{C}|\), as our goal was not to compare entropy values across different agents but rather to assess the effectiveness of configuration entropy as an observable. Specifically, we are interested in whether this measure can serve as a meaningful proxy for the intelligence, in the sense of \textit{skill level}, of an agent.

The results obtained, shown in Figure~\ref{fig:stockfish_entropy}, reveal a clear correlation between configuration entropy and the skill level of Stockfish. In particular, we observe a pronounced discontinuity: while levels \(0\) through \(19\) form a relatively smooth cluster, level \(20\) lies distinctly apart, exhibiting a markedly lower entropy value. This discontinuity is not an artifact of our method, but instead reflects a genuine qualitative difference in the underlying agent. Indeed, only level \(20\) corresponds to the full-strength Stockfish engine, whereas levels \(0\) through \(19\) are artificially weakened variants obtained through the introduction of depotentiation mechanisms. The transition from level \(19\) to level \(20\) is therefore structural, representing the introduction of the unaltered policy, rather than a mere adjustment of parameters.  

It is particularly remarkable that our proposed indicator is sensitive enough to capture this subtle but important distinction, {even with a relatively low number of samples (with respect to the dimension of the phase space) used to estimate the configuration entropy}. The ability of the configuration entropy to reflect not only gradual changes in performance but also sharp, qualitative shifts in agent design strongly supports its interpretation as a meaningful proxy for policy information content. These findings suggest that configuration entropy may serve as a broadly applicable tool for characterizing agent intelligence, even in cases where the underlying policy is not directly accessible.

\section{Conclusions}\label{sec:discuss}

{The concept of random walks has been used in a large variety of different contexts, of broad applied and fundamental interest. Particularly intriguing is the setting where two such walkers are simultaneously evolved on a finite spatial domain, and therein act as a chaser-target pair. Stated differently, each walker explores the surroundings propelled by a self-referred table of stochastic moves, which is not influenced by the opponent position: the game ends when one walker hits the other. Every walker acts therefore as a mobile trap for its competitor. This paradigm can be made more complex by allowing for the walkers to learn from experience. In the chaser-target setting different rewards (linear vs. non linear) can be assigned to break spatial homogeneity. To maximize their income, walkers have to strategically adjust their exploration to favor encounters in the locations that maximize their own reward.  In the setting that we have here analyzed, smart agent learns to hit on the bare counterpart in a region of the one dimensional line where the reward is larger, so as to maximize the earned capital. In a financial trading scenario, as evoked in the introduction, smart buyers (seller) implements a tactical reasoning, tailored to take advantage of the seller (buyer) operation. The goal is to behave in such a way that the encounter (that sets the price of the trading) occurs in the rightmost (leftmost) portion of the directed line of value. 

In this work we have discussed how the reference statistics for the relative encounter of smart walkers can be analytically inferred from the Q-tables of the trained moves. Moreover, and this is the main conclusion, we have shown that configuration entropy can be used as a proxy to gauge the acquired ability of smart operators when facing complex tasks. Configuration entropy, as well as policy entropy, are reported to decay along training, as compared to their initial random reference values. The more complex the task to cope with, the more pronounced the gap dug before asymptotic saturation. The observed gap in the computed configuration entropy can be hence used as a viable measure of the complexity of the task faced by the agent, or as an indirect proxy of the skills acquired upon learning. The identification of a viable proxy for the information gathering of learning agents appears to be particularly useful in real-world contexts (\textit{e.g.} biological), where the reward signals are unknown and the policies are not directly accessible. To challenge the claimed interpretation of entropy as a measure of seemingly rational behavior, we operated the Stockfish chess engine against a quasi-random untrained opponent. Following the analysis, we could report on a statistically meaningful correlation between the computed configuration entropy and the skill level setting of the employed Stockfish engine. 

Summing up, working in the framework of a chaser-target model, with (just one) trainable agent, and leveraging on established reinforcement learning strategies, we proposed and tested the notion of configurational entropy as a possible measure of the rational - non random - operating skills. This is the main message that we intend to deliver. Indeed, the configurational entropy has been validated as viable quantity to measure the level of awareness acquired by Alice upon training. This latter quantity correlates nicely with the policy entropy, which cannot be straightforwardly accessed in real world experiments, as we argued above.  Nicely enough, we show that the configurational entropy can be successfully  applied to a real world challenge (a chess game) for which the policy entropy is not directly available. Configurational entropy is thus a solid measure of agents’ awareness. The investigated toy model (the two walkers sharing a discrete line) served as a  benchmark framework to substantiate our claim, in a controlled setting that makes analytical progress possible (before and after training, in light of the inherent Markovianity of both models).} Extending this analysis to a setting where the walkers are both allowed to learn while exploring a more intricate spatial environment (a lattice in higher dimensions or a heterogeneous graph) constitutes a future direction of possible investigation. For a first step in this direction, see Appendix \ref{app:predator_prey} where two walkers interacting via a \textit{predator-prey} scheme undergo simultaneous training.

\section*{Code availability}

The code regarding the dynamic of walker agents in discrete space is available at the following link: \href{https://github.com/gianluca-peri/smart-walkers}{\nolinkurl{https://github.com/gianluca-peri/smart-walkers}}

\onecolumn

\newpage

\begin{appendices}

\section{Proofs of the closed forms}
\label{app:proofs-of-closed-forms}

\subsection{The encounter probability distribution formula}

In this appendix we present the proof of Eq. \eqref{eq:closed-form-p}. We focus on the probability distribution in the limit of time tending to infinity $\bm{P}_\infty$:
\begin{equation} \label{eq:p_inf}
\bm{P}_\infty = \lim _{t \to +\infty}\tilde{A}^{t-t_{in}} \bm{P}(t_{in}).
\end{equation}
We expand the vector $\bm{P}(t_{in})$ on the basis of {normalized} eigenvectors of $\tilde{A}$. {We denote this basis $\bm{\psi}_\alpha, \text{with} \ \alpha=0,...,N^2-1$.} Thus:
\begin{equation}\label{eq:p_in}
    \bm{P}(t_{in}) = \sum _\alpha \phi_\alpha\bm{\psi} _\alpha,
\end{equation}
with $\phi_\alpha \in \mathbb{R}$.

Substituting Eq. \eqref{eq:p_in} in Eq. \eqref{eq:p_inf}, we obtain:
\begin{equation}
\bm{P}_\infty = \lim _{t \to +\infty}\tilde{A}^{t-t_{in}} \sum _\alpha \phi_\alpha\bm{\psi} _\alpha.
\label{eq:stepping_stone}
\end{equation}
There exists two kinds of eigenvectors for $\tilde{A}$. {The ones associated to the absorbing states that we placed, namely:}
\begin{equation}
    \bm{\psi}_\mu, \ \mu \in \mathcal{B}.
\end{equation}
Notice that such eigenvectors have associated eigenvalues equal to $1$. The second kind of eigenvectors for $\tilde{A}$ are the ones that are associated to eigenvalues smaller than $1$, i.e.:
\begin{equation}
    \bm{\psi} _\nu, \ \nu \in \mathcal{G}.
\end{equation}
Eq. \eqref{eq:stepping_stone} can be written as:
\begin{equation}
\bm{P}_\infty = \lim _{t \to +\infty}\tilde{A}^{t-t_{in}} \left(\sum _{\mu \in \mathcal{B}} \phi _\mu \bm{\psi}_\mu + \sum _{\nu \in \mathcal{G}} \phi _\nu \bm{\psi} _\nu\right).
\label{eq:stepping_stone_after}
\end{equation}
By using the property that  $\bm{\psi}$ are eigenvector of $\tilde{A}$, we obtain:
\begin{equation}\label{eq:eqq}
\bm{P}_\infty = \lim _{t \to +\infty}\left(\sum _{\mu \in \mathcal{B}} \phi _\mu \bm{\psi}_\mu + \sum _{\nu \in \mathcal{G}} \phi _\nu \lambda _\nu^{t-t_{in}}\bm{\psi} _\nu\right).
\end{equation}
In the limit of $t \to \infty$ the second term in r.h.s. of Eq. \eqref{eq:eqq} goes to zero, because $\lambda$'s are less then $1$.

We, therefore, end up with:
\begin{equation}
\bm{P}_\infty = \sum _{\mu \in \mathcal{B}} \phi _\mu \bm{\psi}_\mu.
\end{equation}
Since these eigenvectors correspond to the stationary probability, they have entries equal to \(1\) only in the respective \(\mu\) position, and zero everywhere else.
Since we are only interested in the entries corresponding to the absorbing state (which are the only non-zero ones), we can write:
\begin{equation}
[\bm{P}_\infty]_{\mu'} = \phi _{\mu '}.
\label{eq:almost_done}
\end{equation}
To obtain the probability distribution of first encounters across the sites, it suffices to compute the coefficients of the expansion of the initial state vector $\bm{P}(t_{\text{in}})$ in the eigenvector basis of the transition matrix. These coefficients can be readily determined using the change-of-basis matrix $M$, which maps vectors from the canonical basis to the eigenbasis. Specifically, the vector of coefficients $\bm{\phi}_\mu$, representing $\bm{P}(t_{\text{in}})$ in the eigenbasis, is given by:
\begin{equation}\label{eq:final_app1}
    \bm{\phi} = M\bm{P}(t_{in}) \ \Rightarrow \ \phi _{\mu'} = \sum _\alpha M_{\mu ', \alpha} P(t_{in})_\alpha.
\end{equation}
From Eq. \eqref{eq:almost_done}, we see that this final expression is exactly the result we aimed to prove.

\subsection{The encounter times formula}

To prove \eqref{eq:closed-form-t} we shall first pay attention to the definition of \(T\) in \eqref{eq:t_vector_def}. It is defined from the global transition matrix \(A=A_A \otimes A_B\)
\begin{itemize}
\item putting to zero all the columns corresponding to the encounter indices, so:
\begin{equation}
    i \in \mathcal{M};
    \label{eq:encounter_indices}
\end{equation}
\item transposing the obtained matrix.
\end{itemize}
Given this definition, the mean meeting time starting from the $l$-th configuration ($t_l$), with $l \not= i$, must be:
\begin{equation}
    t_l = \sum _{k=0}^{N^2-1} T_{lk}(t_k + 1),
    \label{eq:start_of_first_proof_1}
\end{equation}
{where leveraged on the standard procedure to calculate expectation values for Markovian systems.}
We can now leverage the normalization of \(T\) to get:
\begin{equation}
    t_l = \sum _{k=0}^{N^2-1} T_{lk}(t_k + 1) = \sum _{k=0}^{N^2-1} T_{lk}t_k +1.
    \label{eq:start_of_first_proof_2}
\end{equation}
In the case \(l=i\) we must have:
\begin{equation}
    t_l = 0 = \sum _{k=0}^{N^2-1} T_{ik}t_k.
    \label{eq:zero_time}
\end{equation}
We can write Eq. \eqref{eq:zero_time} only because we substituted $\bm{O}$, a vector with all zeros, to the columns of \(A\) corresponding to the encounter states. Looking at \eqref{eq:start_of_first_proof_2} and \eqref{eq:zero_time}, we can notice that they can be merged into a single expression:
\begin{equation}
    \bm{t} = T \bm{t} + \bm{c},
    \label{eq:final_of_first_proof}
\end{equation}
where \(\bm{c}\in \mathbb{R}^{N^2}\) must be the vector with \(1\) in all entries except for the ones with index given by \eqref{eq:encounter_indices}, that must be \(0\). 
We can now solve \eqref{eq:final_of_first_proof} to get the final result.

\section{Get the smart transition matrix}
\label{app:get_global_A_from_policy}

To complete the comparison between the smart-agent case (Alice) and the random-agent baseline (Bob), we reconstruct the global time-evolution matrix $A$ directly from the Q-tables of the two walkers. Importantly, because Alice learns a state-dependent policy that accounts for Bob’s position, the resulting matrix $A$ is no longer a priori factorizable into independent components $A_A$ and $A_B$. Instead, the full dynamics must be derived from the Q-tables themselves, denoted $Q_A$ for Alice and $Q_B$ for Bob.

To carry out this reconstruction, it is essential to understand how the tensor product indexes transition probabilities in the joint state space. For two-dimensional matrices, the tensor product corresponds to the Kronecker product, and in the case of two random walkers, the matrix $A$ was initially defined using this construction:

\begin{equation}
    A \ \dot= \ A_A \otimes A_B.
    \label{eq:old_definition}
\end{equation}
From the definition of Kronecker product, this implies:
\begin{equation}
    A_{N \cdot i+i',N \cdot j+j'} = [A_A]_{i,j} \cdot [A_B]_{i',j'},
\end{equation}
and so, from the definition of \(A_A,A_B\), we have:
\begin{equation}
    A_{N \cdot i+i',N \cdot j+j'} = p_A(j \to i) \cdot p_B(j' \to i');
    \label{eq:new_definition}
\end{equation}
in fact recall that the transition matrix element $[A_k]_{ij}, \ k \in \{A,B\}$ has been defined as the probability of transition from state $j$ to state $i$, for the $k$ walker. In the context of random walkers without smart policies, Eq. \eqref{eq:new_definition} is equivalent to Eq. \eqref{eq:old_definition} (in fact, as we just shown, \eqref{eq:new_definition} can be derived from \eqref{eq:old_definition}). For smart walkers, however, the definition of \(A\) in the form \eqref{eq:old_definition} is no longer applicable, since the dynamics is explicitly non factorizable. However, the definition of \(A\) in the form of \eqref{eq:new_definition} is still correct.

We, therefore, adopt \eqref{eq:new_definition} as the definition of the global time evolution matrix \(A\). For smart walkers the probabilities appearing in \eqref{eq:new_definition} are part of their policy \(\pi(a|s)\). In our case, the states contain the position of the agent and the position of the opponent.
We must be careful to avoid mixing up the notation. So we indicate the state with \(m\), which stands for \textit{my position}, and \(o\) stands for \textit{opponent position}. In light of this we have:
\small
\begin{equation}
    \begin{dcases}
        A_{N \cdot i+i',N \cdot j+j'} = 0 & \text{if} \ |i-j|>1 \vee |i'-j'|>1\\
        A_{N \cdot i+i',N \cdot j+j'} = \pi _A \big(-1|m,o\big) \pi _B\big(-1|m',o'\big) & \text{if} \ j = i + 1 \wedge j'=i' + 1\\
        A_{N \cdot i+i',N \cdot j+j'} = \pi _A \big(-1|m,o \big) \pi _B\big(0|m',o'\big) & \text{if} \ j = i + 1 \wedge j'=i'\\
        \cdots\\
        A_{N \cdot i+i',N \cdot j+j'} = \pi _A \big(+1|m,o \big) \pi _B\big(+1|m',o'\big) & \text{if} \ j = i - 1 \wedge j'=i' - 1\\
    \end{dcases}
\end{equation}
\normalsize
where:
\begin{equation}
    m = o' \quad , \quad o = m'.
\end{equation}
Therefore, we have:
\small
\begin{equation}
    \begin{dcases}
        A_{N \cdot i+i',N \cdot j+j'} = 0 & \text{if} \ |i-j|>1 \vee |i'-j'|>1\\
        A_{N \cdot i+i',N \cdot j+j'} = \pi _A \big(-1|m,o\big) \pi _B\big(-1|o,m\big) & \text{if} \ j = i + 1 \wedge j'=i'+1\\
        A_{N \cdot i+i',N \cdot j+j'} = \pi _A \big(-1|m,o\big) \pi _B\big(0|o,m\big) & \text{if} \ j = i + 1 \wedge j'=i'\\
        \cdots\\
        A_{N \cdot i+i',N \cdot j+j'} = \pi _A \big(+1|m,o\big) \pi _B\big(+1|o,m\big) & \text{if} \ j = i - 1 \wedge j'=i'-1\\
    \end{dcases}
\end{equation}
\normalsize
We then notice that \textit{my position} and \textit{opponent position} for the first walker, \textit{in the present}, are simply \(j\) and \(j'\), so:
\small
\begin{equation}
    \begin{dcases}
        A_{N \cdot i+i',N \cdot j+j'} = 0 & \text{if} \ |i-j|>1 \vee |i'-j'|>1\\
        A_{N \cdot i+i',N \cdot j+j'} = \pi _A \big(-1|j,j'\big) \pi _B\big(-1|j',j\big) & \text{if} \ j = i + 1 \wedge j'=i' + 1\\
        A_{N \cdot i+i',N \cdot j+j'} = \pi _A \big(-1|j,j'\big) \pi _B\big(0|j',j\big) & \text{if} \ j = i + 1 \wedge j'=i'\\
        \cdots\\
        A_{N \cdot i+i',N \cdot j+j'} = \pi _A \big(+1|j,j'\big) \pi _B\big(+1|j',j\big) & \text{if} \ j = i - 1 \wedge j'=i'-1\\
    \end{dcases}
\end{equation}
\normalsize
{Rendering the dotted portion explicit here is the final recipe to obtain $A$ from $\pi _A, \pi _B$:
\small
\begin{equation}
    \begin{dcases}
        A_{N \cdot i+i',N \cdot j+j'} = 0 & \text{if} \ |i-j|>1 \vee |i'-j'|>1\\[1em]
        A_{N \cdot i+i',N \cdot j+j'} = \pi _A \big(-1|j,j'\big) \pi _B\big(-1|j',j\big) & \text{if} \ j = i + 1 \wedge j'=i' + 1\\
        A_{N \cdot i+i',N \cdot j+j'} = \pi _A \big(-1|j,j'\big) \pi _B\big(0|j',j\big) & \text{if} \ j = i + 1 \wedge j'=i'\\
        A_{N \cdot i+i',N \cdot j+j'} = \pi _A \big(-1|j,j'\big) \pi _B\big(+1|j',j\big) & \text{if} \ j = i + 1 \wedge j'=i' - 1\\[1em]
        A_{N \cdot i+i',N \cdot j+j'} = \pi _A \big(0|j,j'\big) \pi _B\big(-1|j',j\big) & \text{if} \ j = i \wedge j'=i' + 1\\
        A_{N \cdot i+i',N \cdot j+j'} = \pi _A \big(0|j,j'\big) \pi _B\big(0|j',j\big) & \text{if} \ j = i \wedge j'=i'\\
        A_{N \cdot i+i',N \cdot j+j'} = \pi _A \big(0|j,j'\big) \pi _B\big(+1|j',j\big) & \text{if} \ j = i \wedge j'=i' - 1\\[1em]
        A_{N \cdot i+i',N \cdot j+j'} = \pi _A \big(+1|j,j'\big) \pi _B\big(-1|j',j\big) & \text{if} \ j = i - 1 \wedge j'=i' + 1\\
        A_{N \cdot i+i',N \cdot j+j'} = \pi _A \big(+1|j,j'\big) \pi _B\big(0|j',j\big) & \text{if} \ j = i - 1 \wedge j'=i'\\
        A_{N \cdot i+i',N \cdot j+j'} = \pi _A \big(+1|j,j'\big) \pi _B\big(+1|j',j\big) & \text{if} \ j = i - 1 \wedge j'=i' - 1
    \end{dcases}
    \label{eq:complete_A_from_pis}
\end{equation}
\normalsize 
}

\section{Walkers as predator-prey couples}\label{app:predator_prey}

\begin{figure}[t]
    \centering
    \includegraphics[width=0.6\textwidth]{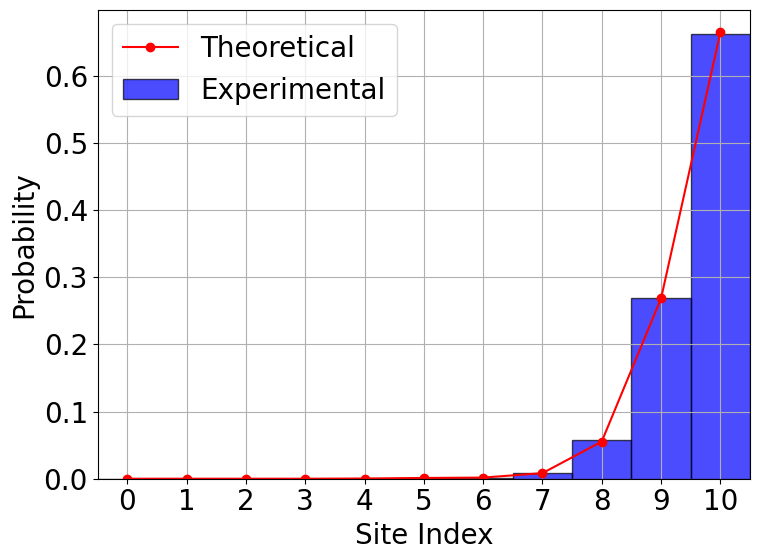}
    \caption{First encounter probability distributions, derived both numerically (\(10{,}000\) simulated games after learning) and analytically via Eq. \eqref{eq:closed-form-p} (note that the softmax temperature $1/\beta$ is fixed at 0.1). Specifically the bar plot reports the different empirical encounter probabilities, while the red curve refers to the computed theoretical predictions.  As for the other setting analyzed throughout the paper, Alice (the predator) starts on the left side, while Bob (the prey) starts on the right. Both walkers converge to the only rational strategy, given the trivial support on which they are both bound: Bob cannot escape the predator since the game board comes to an end at site $10$, but at least he moves all the way to the right to try to delay the encounter, as much as possible. Alice on the other hand has to travel all the way to the right to get the prey. Beyond the game-theoretical interpretation, we remark that the training, as expected, drastically morphs the first encounter probability distribution of the walkers (see Figure \ref{fig:Jacobi}).}
    \label{fig:pppd}
\end{figure}

\begin{figure}[t]
    \centering
    \begin{subfigure}{0.32\textwidth}
        \centering
        \includegraphics[width=\linewidth]{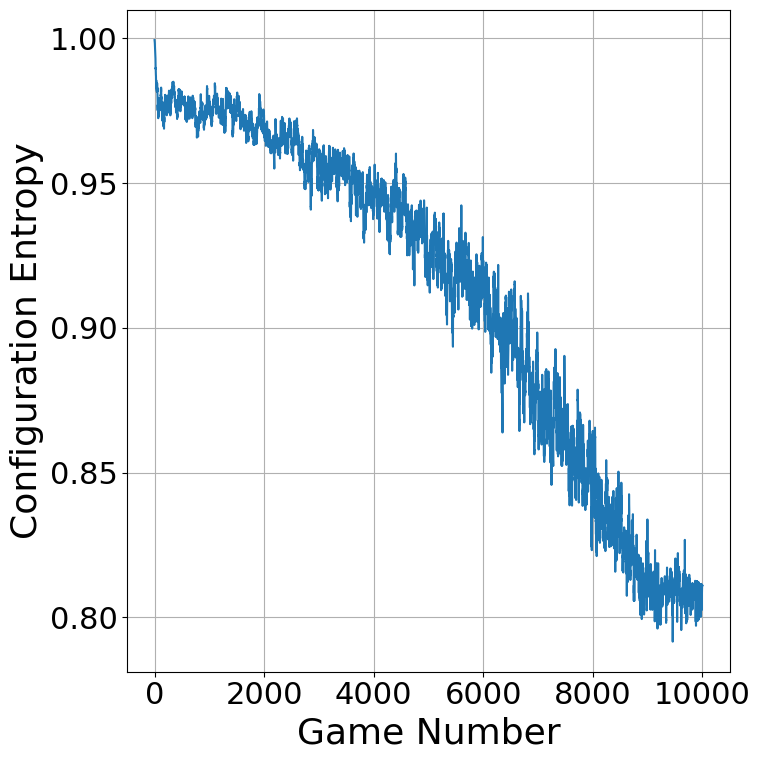}
        \caption{Configuration entropy during training, calculated via \eqref{eq:configuration_entropy}.}
        \label{fig:ppconfent}
    \end{subfigure}
    \hfill
    \begin{subfigure}{0.32\textwidth}
        \centering
        \includegraphics[width=\linewidth]{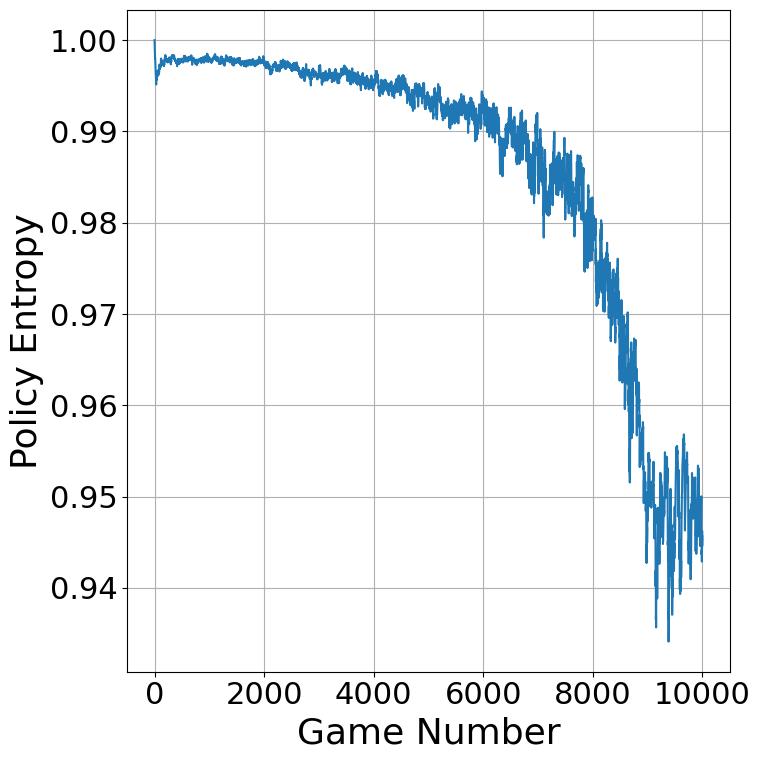}
        \caption{Policy entropy of the predator, Alice (via \eqref{eq:policy_entropy}).}
        \label{fig:ppa}
    \end{subfigure}
    \hfill
    \begin{subfigure}{0.32\textwidth}
        \centering
        \includegraphics[width=\linewidth]{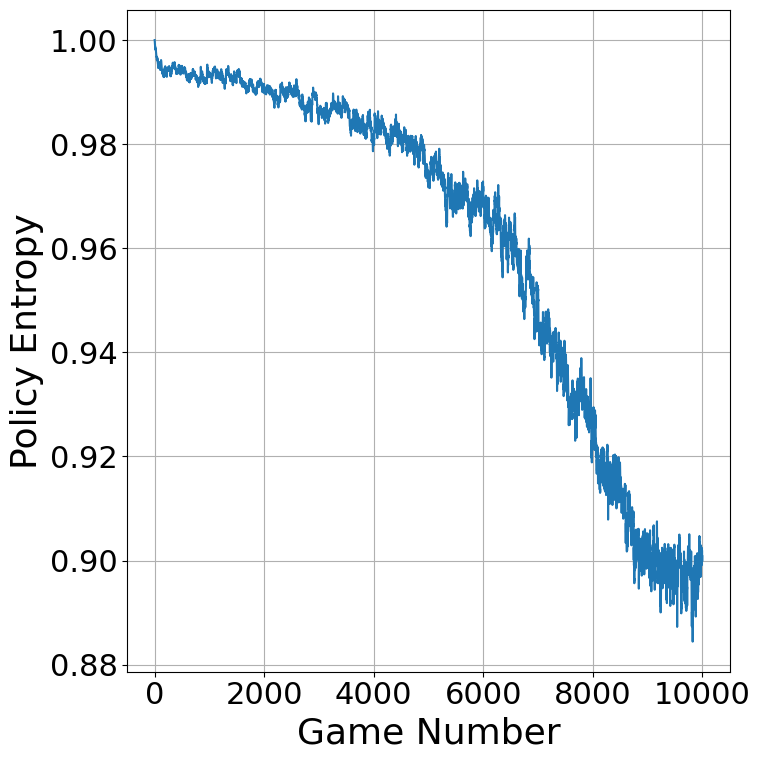}
        \caption{Policy entropy of the prey, Bob (via \eqref{eq:policy_entropy}).}
        \label{fig:ppb}
    \end{subfigure}
    \hfill
    \caption{Different entropies are depicted as measured during the training of Alice (predator) and Bob (prey). Notice that, in this new context where agents are simultaneously trained,  both  entropies (each associated to one of the agents) decrease as a result of the reinforcement learning process. This is also true for the global  configuration entropy of the system, which can be therefore conceptualized as an effective measure of the collective level of acquired awareness.}
    \label{fig:pp_configuration_entropies}
\end{figure}

In the main body of the paper, we focused on scenarios were just one agent undergoes training. As already mentioned above, the formalism here presented applies also to settings were multiple walkers are subjected to simultaneous reinforcement learning. To provide a concrete example, in this Appendix we consider a simplified scenario where two walkers are mutually entangled by \textit{predator-prey} interaction patterns. Specifically, agents are bound to evolve on the very same one dimensional lattice discrete space as employed in the main body of the paper. Therein, we design a reward function that assigns to Alice (the predator) a score $+1$, when an encounter happens. Bob (the prey) is instead given a score $-1$, when hit by Alice. Predator-prey contexts usually impose a time pressure on the agents, and this is accounted for in the analyzed setting by introducing a \textit{discount factor} $\gamma$  smaller than $1$ (specifically, in the following, $\gamma = 0.8$). We also assume that walkers cannot cross each other, as for the other settings here analyzed. We trained both walkers under the above reward pressure:  the resulting (after training) empirical distribution of the first encounter is reported in Figure \ref{fig:pppd}, with the same notation adopted for the similar plots discussed in the main body of the paper. Figure \ref{fig:pp_configuration_entropies}, on the other hand, displays the agents' entropy profiles in this new contexts. Since this time both agents undergo reinforcement learning, as we expect both policies to get optimized away from their initial random walking state, yielding a noticeable drop in the respective policy entropy. Also the configuration entropy of the global system drops during training. It can be hence speculated that the effective drop of this latter quantity provides a sensible measure of the collective intelligent (or non random) behaviour, acquired by experience.

\end{appendices}

\FloatBarrier

\printbibliography

\end{document}